\documentclass[12pt]{extarticle}
\usepackage{geometry}
\usepackage{subfig}
\makeatletter
\newcommand\ackname{Acknowledgements}
\if@titlepage
\newenvironment{acknowledgements}{
	\titlepage
	\null\vfil
	\@beginparpenalty\@lowpenalty
	\begin{center}
		\bfseries \ackname
		\@endparpenalty\@M1
\end{center}}
{\par\vfil\null\endtitlepage}
\else

\usepackage{mathrsfs}
\usepackage{eucal}
\usepackage{amssymb,amsmath}

\usepackage{pgf}
\usepackage{tikz}
\usepackage{cite}
\usepackage{graphicx}
\usepackage{color}
\usepackage{amssymb}
\usepgflibrary{arrows}
\usetikzlibrary{calc,angles,positioning,intersections,quotes,backgrounds,decorations.pathreplacing}
\usetikzlibrary{decorations.markings}
\usepackage{amsmath}
\usepackage{amsthm}
\usepackage{accents}

\theoremstyle{remark}
\theoremstyle{definition}

\numberwithin{equation}{section}

\newcommand{\nn}{\nonumber}
\usepackage{comment}
\usepackage{physics}
\definecolor{mediumtaupe}{rgb}{0.4, 0.3, 0.28}
\definecolor{skobeloff}{rgb}{0.0, 0.48, 0.45}
\definecolor{sandstorm}{rgb}{0.93, 0.84, 0.25}
\definecolor{khaki}{rgb}{0.76, 0.69, 0.57}
\definecolor{olivedrab7}{rgb}{0.24, 0.2, 0.12}
\definecolor{sanddune}{rgb}{0.59, 0.44, 0.09}
\definecolor{mediumseagreen}{rgb}{0.24, 0.7, 0.44}
\definecolor{persianplum}{rgb}{0.44, 0.11, 0.11}

\newcommand{\Lagr}{\mathcal{L}}

\DeclareSymbolFont{rsfs}{U}{rsfs}{m}{n}
\DeclareSymbolFontAlphabet{\mathscrsfs}{rsfs}
\makeatletter
\newcommand*{\rom}[1]{\expandafter\@slowromancap\romannumeral #1@}
\makeatother

\usetikzlibrary{decorations.pathreplacing,decorations.markings}
\tikzset{
	on each segment/.style={
		decorate,
		decoration={
			show path construction,
			moveto code={},
			lineto code={
				\path [#1]
				(\tikzinputsegmentfirst) -- (\tikzinputsegmentlast);
			},
			curveto code={
				\path [#1] (\tikzinputsegmentfirst)
				.. controls
				(\tikzinputsegmentsupporta) and (\tikzinputsegmentsupportb)
				..
				(\tikzinputsegmentlast);
			},
			closepath code={
				\path [#1]
				(\tikzinputsegmentfirst) -- (\tikzinputsegmentlast);
			},
		},
	},
	mid arrow/.style={postaction={decorate,decoration={
				markings,
				mark=at position .5 with {\arrow[#1]{stealth}}
	}}},
}
\begin{document}
	\title{\textbf{The multi-time propagators and the consistency condition}}
	\author{Siwaporn Sungted and  Sikarin Yoo-Kong \\
		\small {The Institute for Fundamental Study (IF),} \\ \small\emph{Naresuan University, Phitsanulok, Thailand, 65000.}\\
	}
	\date{}
	\maketitle
	\abstract
	For a non-relativistic quantum system of $N$ particles, the wave function is a function of $3N$ spatial coordinates and one temporal coordinate. The relativistic generalisation of this wave function is a function of $N$ time variables known as the multi-time wave function and its evolution is described by $N$ Schr\"{o}dinger equations, one for each time variable. To guarantee the existence of a non-trivial common solution to these $N$ equations, the $N$ Hamiltonians need to satisfy a compatible condition known as an integrability condition. In this work, the integrability condition will be expressed in terms of Lagrangians. The time evolution of a wave function with $N$ time variables through the Feynman picture of quantum mechanics is derived. However, these evolutions will be compatible if and only if the $N$ Lagrangians satisfy a certain relation called the consistency condition which could be expressed in terms of the Wilson line. As a consequence of this consistency condition, the evolution of the wave function gives rise to a key feature called the ``path-independent" property on the space of time variables. This would suggest that one must consider all possible paths not only on the space of dependent variables(spatial variables) but also on the space of independent variables(temporal variables). In the view of the geometry, this consistency condition can be considered as a zero curvature condition and the multi-time evolutions can be treated as compatible parallel transport processes on flat space of time variables.
	\section{Introduction}
	In non-relativistic quantum mechanics, the wave function for $N$ particles can be expressed as $\Psi(q_1,q_2,...,q_N,t)$, where $q_k\in \mathbb R^d$, $k=1,2,...,N$. If one asks for the relativistic counterpart of this wave function we encounter with the difficulty as follows. Since there is only one time variable in the wave function, it is not clear how one could perform the Lorentz transformation. The argument of $\Psi$ can be treated as a collection of $N$ simultaneous space-time points $(t,q_1),...,(t,q_N)$ which under the Lorentz transformation is changed to $(t'_1,q'_1),...,(t'_N,q'_N)$, of course, in general, $t'_1\neq t'_2\neq...\neq t'_N$. Then it is quite natural to introduce the multi-time structure into the wave function $\Phi(q_1,t_1,...,q_N,t_N)$ to manifest the Lorentz transformation. This idea was first introduced by Dirac in 1932 \cite{PAM}. Consequently, we have a set of partial differential equations \footnote{Here we set $\hbar=1$ throughout the text.}
	\[
	i\pdv{}{t_j}\Phi(q_1,t_1,...,q_N,t_N)=H_j\Phi(q_1,t_1,...,q_N,t_N)\;,
	\]
	where $j=1,2,...,N$ and $H_j$ is a self-adjoint Hamiltonian for the $j^{\text {th}}$ particle. These multi-time systems will be compatible or a common non-trivial solution $\Phi$ exists if and only if the relation 
	\textcolor{black}{
	\[
	\pdv{H_{j}}{t_{k}}-\pdv{H_{k}}{t_{j}}-i\left[H_{j},H_{k}\right]=0\;,\;\;\forall j\neq k \;
	\]
	}
	hold. This is known as the consistency condition or integrability condition \cite{Nickel,Deckert,Petratthesis,Petrat,PetratCandQ}.
	\\
	\\
	The idea of the multi-time wave function formalism could be possibly useful in many aspects. For example, Petrat and Tumulku \cite{PetratQFT} demonstrated that the relevant interacting quantum field theories can be reformulated in terms of multi-time wave functions and therefore, multi-time wave function, the Tomonaga-Schwinger and the Heisenberg approaches are equivalent and the consistency condition of the multi-time formulation explains why in nature the process that a fermion decays into two fermions cannot happen \cite{PetratPairCreation}. Lienert, Petrat and Tumulka \cite{Lienert} pointed out that multi-time wave function can be considered in discrete action principles and can be applied to study the cellular automata.
	\\
	\\
	Here comes the main question of this work. \emph{What is the Lagrangian analogue of the consistency condition?} This question is natural to be asked since normally in physics we could choose to work with Hamiltonian or Lagrangian descriptions. Then in this work, the variational principle will play a central role in order to obtain the consistency condition or integrability condition and the quantum multi-time evolution will be captured through Feynman's path integration expressing in terms of the Wilson line. To make things flowing smoothly, the remaining body of this paper is organised as follows. In section \ref{section2}, a brief review of the Hamiltonian approach both classical and quantum levels will be given. In section \ref{section3}, the derivation on integrability condition through the variational principle will be explained in the classical case. After that, the multi-time propagators will be constructed and quantum multi-time evolution is studied. The conclusion will be given in the last section.

	\section{Hamiltonian approach}\label{section2}
	\subsection{Classical case}
	In this section, we will give a short review on multi-time structure in the context of classical mechanics as well as the derivation of the consistency criterion known as the Hamiltonian commuting condition \cite{Longhi,PetratCandQ}. 
	\\
	\\
	Suppose there are a set of Hamiltonians $\{H_1,H_2,...,H_N \}$ and a multi-time Hamilton-Jacobi function $\Phi$ associated with a set of time variables $t=(t_1,t_2,...,t_N)$, where $t_j\in \mathbb R$ . We then look for solutions for a set of the first order differential equations given by
	\begin{eqnarray}
		\frac{\partial }{\partial t_j}\Phi( q, t)+H_j\left( q, t,\frac{\partial}{\partial  q}\Phi( q, t)\right)=0\;,\;\;\text{where}\; q\in \mathbb R^d \;\text{and}\; j=1,2,3,...,N
		\;. \label{E1}
	\end{eqnarray}
	It is well known that a set of equations in \eqref{E1} is the multi-time Hamilton-Jacobi equations and is overdetermined. Then, to get a nontrivial common solution, one may need all Hamiltonians to commute in an appropriate way known as the Hamiltonian commuting flows. To obtain that particular consistency condition, we look at the compatible flows between $t_i$ and $t_j$. What we have now are 
	\begin{align}
		\frac{\partial^2}{\partial{t_{j}}\partial{t_{i}}}\Phi=-\pdv{H_{i}}{t_{j}}+\pdv{H_{i}}{\left(\frac{\partial\Phi}{\partial q_k}\right)}\left(\pdv{H_{j}}{ q_k}+\pdv{H_{j}}{\left(\frac{\partial\Phi}{\partial q_l}\right)}\centerdot\frac{\partial^2{}}{\partial{ q_k}\partial{ q_l}}\Phi\right) \;, \label{CHJ}
	\end{align}
	and
	\begin{align}
		\frac{\partial^2}{\partial{t_{i}}\partial{t_{j}}}\Phi=-\pdv{H_{j}}{t_{i}}+\pdv{H_{j}}{\left(\frac{\partial\Phi}{\partial q_k}\right)}\left(\pdv{H_{i}}{ q_k}+\pdv{H_{i}}{\left(\frac{\partial\Phi}{\partial q_l}\right)}\centerdot\frac{\partial^2{}}{\partial{ q_k}\partial{ q_l}}\Phi\right) \;. \label{CHJ}
	\end{align}
	The compatibility requires
	\begin{equation}
		\left(\frac{\partial^2{}}{\partial{t_{j}}\partial{t_{i}}}-\frac{\partial^2{}}{\partial{t_{i}}\partial{t_{j}}}\right)\Phi=0\nn \;,
	\end{equation}
	leading to the condition
	\begin{align}
		-\pdv{H_{i}}{t_{j}}+\pdv{H_{j}}{t_{i}}-\left\{H_{i},H_{j}\right\}=0 \;, \label{ccc}
	\end{align}
	where $\{ . , .\}$ is the standard Poisson bracket.
	%
	\subsection{Quantum case}
	\textbf{Single-time case}: A natural way to move from the classical level to the quantum level through the Hamiltonian function is the Schr\"{o}dinger approach. Given a state $\Psi(q,t)$ defined in the Hilbert space $\mathscr H$, 
	the Schr\"{o}dinger equation of a particle with mass $m$ trapped in the potential $V$ is
	\begin{equation}
		i\pdv{\Psi}{t}=H\Psi\;,\label{e1}
	\end{equation}
	where $H$ is the Hamiltonian operator. 
	%
	%
	The time evolution of the wave function can be considered through the unitary operator $U(t',t)$, where $t'>t$ such that 
	\begin{equation}
		\Psi(q',t' )=U(t',t)\Psi(q,t)\;, \label{eq:7}
	\end{equation}
	where
	\begin{equation}
		U(t',t)=e^{-i\int_t^{t'}{H}(\tau)d\tau}\;.
	\end{equation}
	It might happen that the Hamiltonian operators evaluated at different moments of time do not commute, i.e., $[H(s),H(s')]\neq 0$. Then, in this situation, the time evolution operator becomes
	\begin{align}
		U(t,s)&=\mathsf T e^{-i\int_{s}^{t}dTH(T,s)} \nonumber \\
		&=I+ \sum_{n=1}^{\infty}(-i)^{n}\int_{s}^{t}dT_{1}\int_{s}^{T_{1}}dT_{2}...\int_{s}^{T_{n-1}}dT_{n}H(T_{1},s)H(T_{2},s)...H(T_{n},s)\;, \label{dyson}
	\end{align}
	where $\textsf{T}$ is the time ordering operator and this expansion is known as the Dyson series \cite{Sakurai}.\\\\
	However, in this study, we restrict ourselves in the case of time-independent Hamiltonians.
	\\
	\\
	\textbf{Multi-time case}: 
	In this case, suppose there are $N$ particles in the system and $(q_1,q_2,..,q_N)$ is a set of coordinates. The single-time wave function is given by $\Psi(q_1,q_2,...,q_N,t)$ and the relativistic version is $\Phi(q_1,t_1,q_2,t_2,...,q_N,t_N)$ satisfying $N$ separable time-dependent Schr\"{o}dinger equations \cite{PAM,Tomonaga}
	\begin{equation}
		\Bigg(H_{j}+\frac{1}{i}\pdv{}{t_{j}}\Bigg)\Phi(q_{1},t_{1},q_{2},t_{2},...,q_{N},t_{N}) = 0\;,\;\;j=1,2,...,N\;,\label{eq:21}
	\end{equation}
	where $H_j$ are the free Schr\"{o}dinger Hamiltonians (or free Dirac Hamiltonians). The ordinary probability amplitude is retrieved by setting all-time coordinates equal
	\begin{equation}
		\Phi(q_1,t,q_2,t,...,q_N,t)=\Psi(q_1,q_2,...,q_N,t)\;.\label{T2}
	\end{equation}
	Here the single-time wave function $\Psi$ satisfies the standard Schr\"{o}dinger equation \eqref{e1} and
	$H = \sum_{j=1}^{N} H_{j}$. Equations \eqref{T2} and \eqref{e1} suggest that the multi-time wave function coincides with the single-time wave function with respect to the Lorentz frame on configurations of $N$ space-time points \cite{Lienert}.
	\\\\
	Here comes to an interesting feature of the system of equations \eqref{eq:21}. The multi-time evolution must satisfy a certain condition. Suppose the multi-time wave function evolves from the initial point $(0,0)$ to the final point $(t_{1},t_{2})$\footnote{For simplicity, we consider only two-time variables.}. In the case of time-independent Hamiltonians, we define $U_1(t_1)=e^{-iH_1t_1}$ as the unitary time operator in $t_1$ direction and $U_2(t_2)=e^{-iH_2t_2}$ as the unitary time operator in $t_2$ direction. There are two different ways to proceed the evolution map as follows
	\begin{figure}[h]
		\centering
		\begin{tikzpicture}
			
			\foreach \coord/\label [count=\xi] in {
				{0.05,0.35}/{$\Phi(0, 0)$},
				{3.95,2.65}/{$\Phi(t_{1}, t_{2})$},
				{3.95,0.35}/{$\Phi(t_{1}, 0)$},
				{0.05,2.65}/{$\Phi(0, t_{2})$}
			}{
				\pgfmathsetmacro\anch{mod(\xi,2) ? "north" : "south"}
				\node[anchor=\anch] at (\coord) {\label};
			}
			
			\draw[-stealth,line width=1pt] (0.8,0) -- (3.1,0) node[midway,below] {$U_1$};
			\draw[-stealth,line width=1pt] (4,0.3) -- (4,2.7) node[midway,right] {$U_2$};
			\draw[-stealth,line width=1pt,dashed,] (0,0.3) -- (0,2.7) node[midway,left] {$U_2$};
			\draw[-stealth,line width=1pt,dashed,] (0.8,3) -- (3.1,3) node[midway,above] {$U_1$};
		\end{tikzpicture}
		\caption{Two compatible maps of the wave function from the initial point $(0,0)$ to the final point $(t_{1},t_{2})$. 
		}
		\label{fig:path}
	\end{figure}
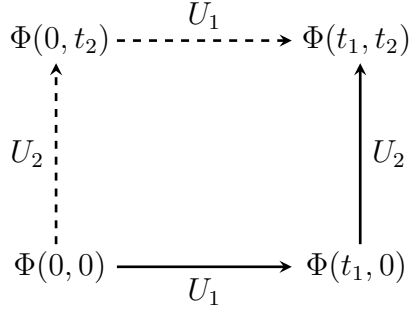

	\begin{equation}
		\Phi(t_{1},t_{2})=e^{-iH_{2}t_{2}}\Phi(t_{1},0)=e^{-iH_{2}t_{2}}e^{-iH_{1}t_{1}}\Phi(0,0)=U_2U_1\Phi(0,0)\label{EC1}
	\end{equation}
	and
	\begin{equation}
		\Phi(t_{1},t_{2})=e^{-iH_{1}t_{1}}\Phi(0,t_{2})=e^{-iH_{1}t_{1}}e^{-iH_{2}t_{2}}\Phi(0,0)=U_1U_2\Phi(0,0)\;.\label{EC2}
	\end{equation}
	From the equations \eqref{EC1} and \eqref{EC2}, the evolution is compatible if and only if
	\begin{equation}
		\left[H_{1},H_{2}\right]=0\;, \label{eq:35}
	\end{equation} 
	which is called the consistency condition or integrability criterion for the multi-time evolution, see figure \ref{fig:path}. 
	In the case of the time-dependent Hamiltonian, one could obtain the consistency condition as \cite{Petrat}
	\textcolor{black}{
	\begin{align}
		\pdv{H_{j}}{t_{k}}-\pdv{H_{k}}{t_{j}}-i\left[H_{j},H_{k}\right]=0\;,\;\;\forall j\neq k \;. \label{qcc}
	\end{align}
	}
	Here equation \eqref{qcc} can be considered as the quantum analogue of the equation \eqref{ccc}. 
	\\
	\\
	\emph{Remark 1}: The wave function $\Phi$ is defined only on the space-like configurations. For a fixed number of particles, the system of multi-time equations with interaction potentials automatically violates the consistency condition \cite{Petrat}. Nevertheless, there is a special initial datum, i.e. setting all time variables equal to zero. In spite of inconsistency, a system of $N$ Schr\"{o}dinger equation can be simultaneously solved \cite{Lill}.
	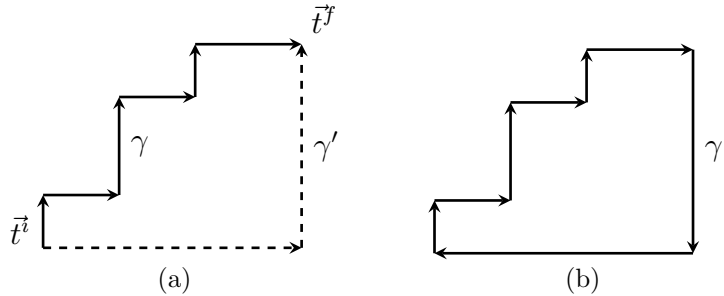
\begin{figure}[h]
		\centering
		\subfloat[]{\label{fig:minipathA}
			\begin{tikzpicture}
				
				\foreach \coord/\label [count=\xi] in {
					{-0.3,0.57}/{$\vec{t}^{i}$},
					{3.7,2.7}/{$\vec{t}^{f}$} 		
				}{
					\pgfmathsetmacro\anch{mod(\xi,2) ? "north" : "south"}
					\node[anchor=\anch] at (\coord) {\label};
				}
				
				\draw[-stealth,line width=1pt] (0,0) -- (0,0.7);
				\draw[-stealth,line width=1pt] (0,0.7) -- (1,0.7);
				\draw[-stealth,line width=1pt] (1,0.7) -- (1,2) node[midway,right] {$\gamma$};
				\draw[-stealth,line width=1pt] (1,2) -- (2,2);
				\draw[-stealth,line width=1pt] (2,2) -- (2,2.7);
				\draw[-stealth,line width=1pt] (2,2.7) -- (3.4,2.7);
				\draw[-stealth,line width=1pt,dashed] (0,0) -- (3.4,0);
				\draw[-stealth,line width=1pt,dashed] (3.4,0) -- (3.4,2.7) node[midway,right] {$\gamma'$};
			\end{tikzpicture}
		}
		\qquad
		\subfloat[]{\label{fig:minipathB}
			\begin{tikzpicture}
				
				\draw[-stealth,line width=1pt] (0,0) -- (0,0.7);
				\draw[-stealth,line width=1pt] (0,0.7) -- (1,0.7);
				\draw[-stealth,line width=1pt] (1,0.7) -- (1,2);
				\draw[-stealth,line width=1pt] (1,2) -- (2,2);
				\draw[-stealth,line width=1pt] (2,2) -- (2,2.7);
				\draw[-stealth,line width=1pt] (2,2.7) -- (3.4,2.7);
				\draw[-stealth,line width=1pt] (3.4,0) -- (0,0);
				\draw[-stealth,line width=1pt] (3.4,2.7) -- (3.4,0) node[midway,right] {$\gamma$};
			\end{tikzpicture}
		}
		\caption{(a) Two different paths $\gamma$ and $\gamma'$ from the initial point $\vec{t}^{i}$ to the final point $\vec{t}^{f}$. (b) A loop $\gamma$.}
		\label{fig:minipath}
	\end{figure}
	\\
	The condition \eqref{qcc} implies the path-independent feature of the time evolution in the context of multi-time quantum theory. This can be seen by the following construction. If we consider the path which is parametised by $\gamma$, see figure \ref{fig:minipathA}, where $\gamma : [0,1]$ from the initial point $\gamma(0)=\vec{t}^{i}=({t_{1}}^{i},{t_{2}}^{i},...,{t_{N}}^{i})$ to the final point $\gamma(1)=\vec{t}^{f}=({t_{1}}^{f},{t_{2}}^{f},...,{t_{N}}^{f})$, the time evolution operator along this particular path is given by
	\begin{equation}
		U_{\gamma} = \mathsf T e^{-i\int_{\gamma}\sum_{j}H_{j}dt_{j}}\;.
	\end{equation}
	Another path parametised by $\gamma'$, see also figure \ref{fig:minipathA}, where $\gamma' : [0,1]$ from the initial point $\gamma'(0)=\vec{t}^{i}=({t_{1}}^{i},{t_{2}}^{i},...,{t_{N}}^{i})$ to the final point $\gamma'(1)=\vec{t}^{f}=({t_{1}}^{f},{t_{2}}^{f},...,{t_{N}}^{f})$, the time evolution operator along this path is given by
	\begin{equation}
		U_{\gamma'} = \mathsf T e^{-i\int_{\gamma'}\sum_{j}H_{j}dt_{j}}\;.
	\end{equation}
	The path-independent feature requires the condition $U_\gamma= U_{\gamma'}$.
	\\
	\\
	In the language of geometry, we can put the path-independent feature as the parallel transport process. To see this, we define the covariant derivative
	\begin{equation}
		\bigtriangledown_{j} = \partial_{j}-iA_{j}\;,
	\end{equation}
	where $\partial_{j}=\partial/\partial_{t_{j}}$ and connection coefficient $A_{j}=-H_{j}$. Then $U_\gamma$ can be treated as the parallel transport operator along the path $\gamma$ known as the order path integral or Wilson line. For any arbitrary loop $\gamma$, see figure \ref{fig:minipathB}, one can express the transport operator in the form 
	\begin{equation}
		U_{\gamma} =\mathsf T e^{-i\oint_{\gamma}\sum_{j}H_{j}dt_{j}}\;,
	\end{equation}
	which is known as the Wilson loop.
	Then the path-independent property is nothing but saying that all closed paths $\gamma$ have trivial holonomy, i.e., $U_{\gamma}=I$. Consequently, a gauge connection processes trivial holonomies if and only if its curvature $F$ defining as
	\begin{equation}
		F_{jk} \equiv -\pdv{H_{k}}{t_{j}}+\pdv{H_{j}}{t_{k}}-i\big[H_{j},H_{k}\big]
	\end{equation}
	vanishes \cite{Petrat}:
	\begin{equation}
		F_{jk}=0 \,\,\,\,\,\,\, \forall j \neq k\;.\label{F1}
	\end{equation}
	With the definition of the curvature, we can rewrite the argument of the exponential of the time evolution operator as
	\begin{equation}
		-i\oint_{\partial\Sigma}\sum_{j}H_{j}dt_{j}=-i\iint_{\Sigma}\sum_{ij}F_{ij}dt_i\wedge dt_{j}\;,\label{PPP}
	\end{equation}
	where $\Sigma$ is a 2-dimensional surface whose boundary is $\partial\Sigma$.
	Obviously, condition \eqref{F1} is identical to \eqref{qcc} so we can consider the consistency condition in the viewpoint of curvature. We knew that curvature is the tool to test the difference of vector that parallel transported along a closed path. If the direction of the initial and the final vector is not different, there is no curvature of the surface, $F_{jk}=0$, which means flat surface. Therefore, the consistency condition \eqref{qcc} of the multi-time wave function can be treated as the zero curvature condition.
	\section{Lagrangian approach}\label{section3}
	In the previous section, the consistency conditions for multi-time evolution both classical and quantum levels are captured through the Hamiltonian picture. Here, in this section, we will express the consistency condition, both classical and quantum cases, in terms of the Lagrangian.
	\subsection{Classical case}
\begin{figure}[h]
		\centering
		\begin{tikzpicture}[x=1cm, y=1cm, z=-0.6cm, scale=0.87]
			\foreach \coord/\label [count=\xi] in {
				{1.8,3.6}/{$q(t_{1}(\sigma_{0}),t_{2}(\sigma_{0}))$},
				{4.6,1.4}/{$q(t_{1}(\sigma_{1}),t_{2}(\sigma_{1}))$},	
				{-1.37,-0.05}/{$(t_{1}(\sigma_{0}),t_{2}(\sigma_{0}))$},
				{4.5,-2}/{$(t_{1}(\sigma_{1}),t_{2}(\sigma_{1}))$},
				{3,-0.3,1}/{$\Gamma$},
				{1.4,-1.17,0.07}/{$\delta t$},
				{2.7,-0.1,2.5}/{$\Gamma'$}
			}{
				\pgfmathsetmacro\anch{mod(\xi,2) ? "north" : "south"}
				\node[anchor=\anch] at (\coord) {\label};
			}
			
			\draw [-stealth,line width=1pt] (0,0,0) -- (4,0,0) node [right] {$t_{1}$};
			\draw [-stealth,line width=1pt] (0,0,0) -- (0,4,0) node [left] {$q$};
			\draw [-stealth,line width=1pt] (0,0,0) -- (0,0,3) node [left] {$t_{2}$};
			Dashed lines
			\draw [loosely dashed]
			(0.5,3,0.5) -- (0.5,0,0.5)
			(4.5,3,2.5) -- (4.5,0,2.5)
			;
			
			\path [draw,postaction={on each segment={mid arrow}},line width=1.25pt]
			(0.5,3,0.5) to [bend left=45] (2.5,3,1.5)
			(2.5,3,1.5) to [bend right=45] (4.5,3,2.5)
			(0.5,0,0.5) to [bend left=45] (2.5,0,1.5)
			(2.5,0,1.5) to [bend right=45] (4.5,0,2.5)
			;
			
			\path [draw,postaction={on each segment={mid arrow}},line width=1.25pt,dashed]
			(0.5,0,0.5) to [bend left=45] (2.6,0,2.3)
			(2.6,0,2.3) to [bend right=60] (4.5,0,2.5)
			;
			
			\draw[-stealth,line width=1pt] (1.8,-1,0.15) -- (1.5,-1,0.5);
			
		\end{tikzpicture}
		\caption{The variation of the path on the space of two-time variables.}
		\label{fig:changevariblesofs}
	\end{figure}
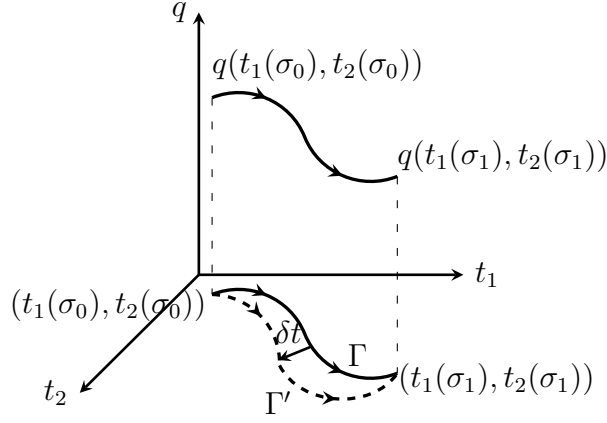

	We start to give the action functional along path $\Gamma$, see figure \ref{fig:changevariblesofs} in the case of two-time variables, defined on the space of time variables
	\begin{align}
		S_{\Gamma}[t]=\int_{\Gamma}\sum_{i=1}^{N} L_{i}dt_{i}
		\;, \label{cptblt2}
	\end{align}
	where \textcolor{black}{$L_{i}=L_{i}(dq_i/dt_i, q_{i}; t)$} is the Lagrangian for $i^{th}$ particle. We introduce a new variable $\sigma_{0}\leqslant\sigma\leqslant\sigma_{1}$ such that $(t_{1}(\sigma),t_{2}(\sigma),...,t_{N}(\sigma))$. Then the action \eqref{cptblt2} becomes
	\begin{align}
		S_{\Gamma}[t(\sigma)]=\int_{\sigma_{0}}^{\sigma_{1}}\Lagr d\sigma \;, \;\;\mbox{where}\;\; \Lagr= \sum_{i=1}^{N}L_{i}\frac{dt_{i}}{d\sigma}\;.
		\label{cptblt2.5}
	\end{align}
	In order to capture the consistency condition for multi-time evolution, we consider the time variation $t_{i}\to t_{i}+\delta t_{i}$ resulting in a new path $\Gamma'$ with the action
	\begin{align}
		S_{\Gamma'}\left[t(\sigma)+\delta t(\sigma)\right]=\int_{\sigma_{0}}^{\sigma_{1}}d\sigma\left( \sum_{i=1}^{N} L_{i}\left(t+\delta t\right)\frac{d\left(t_{i}+\delta t_{i}\right)}{d\sigma}\right)
		\;. \label{cptblt3}
	\end{align}
	\\
	Employing the Taylor series expansion and ignoring higher-order terms, therefore each Lagrangian can be expressed as
	\begin{align}
		L_{i}\left(t+\delta t\right)=L_{i}\left(t\right)+\sum_{j=1}^{N} \delta t_{j}\pdv{L_{i}}{t_{j}}+... \;, \;\;i=1,2,...,N\;. \label{taylorL}
	\end{align}
	The variation of the action is given by
	\begin{align}
		S_{\Gamma'}\left[t(\sigma)+\delta t(\sigma)\right]-S_{\Gamma}[t] \equiv \delta S \approx
		\int_{\sigma_{0}}^{\sigma_{1}}d\sigma
		\left\{\sum_{i=1}^{N} \left( \sum_{j=1}^{N} \delta t_{j} \pdv{L_{i}}{t_{j}}\right) \frac{dt_{i}}{d\sigma} + \sum_{i=1}^{N}L_{i}\frac{d \delta t_{i}}{d\sigma}\right\}
		\;. \label{cptblt4}
	\end{align}
	Using integration by parts, \eqref{cptblt4} becomes
	\begin{align}
		\delta S&=\int_{\sigma_{0}}^{\sigma_{1}}d\sigma
		\left\{ \sum_{i,j=1}^{N}
		\delta t_{i}\left(\pdv{L_{j}}{t_{i}}-\pdv{L_{i}}{t_{j}}\right)\frac{dt_{j}}{d\sigma}\right\}
		\;,\;\;\forall i\neq j \;. \label{cptblt6}
	\end{align}
	Imposing the condition $\delta S =0$, we obtain
	\begin{align}
		\pdv{L_{j}}{t_{i}}=\pdv{L_{i}}{t_{j}}
		\;,\;\;\forall i\neq j \;. \label{cptblt7}
	\end{align}
	Equation \eqref{cptblt7}\footnote{This equation was first derived in a different context, the integrable 1-dimensional many-body system \cite{SY}, to capture also the consistency condition.} is nothing but the consistency condition for the multi-time evolution in terms of the Lagrangian. Consequently, under condition \eqref{cptblt7} the action remains the same under the path variation on the space of time variables. This is nothing but the path-independent feature of the evolution on the space of time variables.
	\\
	\\
	\emph{Remark 2}: We shall point out that one can do the variation on the action with respect to the coordinate variables resulting in a set of Euler-Lagrange equations together with constraints \cite{SY}.
	\\
	\\
	From the geometric point of view, equation \eqref{cptblt7} can also be obtained. Suppose that $\alpha$ is a differential ($k$-1)-form. The generalised Stokes' theorem states that \textit{the integral of its exterior derivative over the surface of smooth oriented k-dimensional manifold $\Sigma$ is equal to its integral of along the boundary $\partial\Sigma$ of the manifold $\Sigma$ }\cite{Fortney}:
	\begin{eqnarray}
		\int_{\partial\Sigma} \alpha = \iint_\Sigma d\alpha \;. \label{eq1}
	\end{eqnarray}
	We now introduce an object $dS$ given by 
	\begin{eqnarray}
		d{{S}}= \;\sum_{i=1}^{N}L_{i}dt_i \;,
	\end{eqnarray}
	as a 1-form on the $N$-dimensional space of independent variables and, therefore, the action \eqref{cptblt2} becomes $S=\int_\Gamma dS$. Applying an exterior derivative to the smooth function coefficients which, in this case, is the Lagrangian, (\ref{eq1}) becomes
	\begin{eqnarray}
		\begin{aligned}
			\oint_{\partial\Sigma}\sum_{i=1}^{N}L_{i}dt_i = & \iint_{\Sigma} \sum_{1\leq i < j \leq N}^{N} \bigg( \frac{\partial L_j}{\partial t_i}-\frac{\partial 
				L_i}{\partial t_j}\bigg) dt_i\wedge dt_j\;.
		\end{aligned}\label{eq2}
	\end{eqnarray}
	The left-hand side of \eqref{eq2} is equivalent to $\int_\Gamma dS-\int_{\Gamma'} dS$.
	Thus, the right-hand side of \eqref{eq2} vanishes, since the exterior derivative operating on the closed-form gives a vanishing result. Therefore, we obtain
	\begin{eqnarray}
		\frac{\partial L_j}{\partial t_i}-\frac{\partial 
			L_i}{\partial t_j} = 0 \; , \quad i, j = 1, 2, 3, ..., N \quad \mbox{and} \quad i\neq j\;,\label{fff}
	\end{eqnarray}
	which are the consistency conditions of the system that evolves in the $N$-dimensional space of independent variables. The main point is that equation \eqref{eq2} is the Lagrangian version of parallel transport feature, see equation \eqref{PPP}, if one defines 
	\begin{eqnarray}
		F_{ij}=\frac{\partial L_j}{\partial t_i}-\frac{\partial 
			L_i}{\partial t_j} \; , \quad i, j = 1, 2, 3, ..., N \quad \mbox{and} \quad i\neq j\;.
	\end{eqnarray}
	Consequently, the consistency condition \eqref{fff} of multi-time evolution can be treated as the zero curvature condition in terms of the Lagrangians.
	\\
	\\
	We find that the condition \eqref{cptblt7} violates if there is the interaction. To see this, we give a simple example as follows. Given 
	%
	$L_{1}=\frac{m\dot{q_{1}}^{2}}{2}+kq_{1}q_{2} $ and
	$ L_{2}=\frac{m\dot{q_{2}}^{2}}{2}$
	\;, \label{checkL}
	%
	where \textcolor{black}{$q_1=q_1(t_1)$, $q_2=q_2(t_2)$}, $k$ is the constant and then 
	\begin{align}
		\pdv{L_{1}}{t_{2}}&=kq_{1}\pdv{q_{2}}{t_{2}} \;,\\
		\pdv{L_{2}}{t_{1}}&=0
		\;. \label{checkL1L2}
	\end{align}
	Thus the interaction leads to inconsistency. We shall see later in the quantum case that the interaction gives also incompatible quantum evolution in terms of the propagators.
	
	\subsection{Quantum case}
\begin{figure}[h]
		\centering
		\begin{tikzpicture}
			
			\foreach \coord/\label [count=\xi] in {
				{0,0}/{$(q_{1},q_{2},t_{1},t_{2})$},
				{4,3}/{$(q'_{1},q'_{2},t'_{1},t'_{2})$},
				{4,0}/{$(\tilde{q}_{1},\tilde{q}_{2},t'_{1},t_{2})$},
				{0,3}/{$(\bar{q}_{1},\bar{q}_{2},t_{1},t'_{2})$},
				{0.5,2.75}/{$\gamma'$},
				{3.5,0.25}/{$\gamma$}
			}{
				\pgfmathsetmacro\anch{mod(\xi,2) ? "north" : "south"}
				\node[anchor=\anch] at (\coord) {\label};
			}
			
			\draw[line width=1pt,postaction={on each segment={mid arrow}} ] (0,0) -- (4,0);
			\draw[line width=1pt,postaction={on each segment={mid arrow}} ] (4,0) -- (4,3);
			\draw[line width=1pt,dashed,postaction={on each segment={mid arrow}} ] (0,3) -- (4,3);
			\draw[line width=1pt,dashed,postaction={on each segment={mid arrow}} ] (0,0) -- (0,3);
		\end{tikzpicture}
		\caption{Two different paths $\gamma$ and $\gamma'$ from the initial point $(t_{1},t_{2})$ to the final point $(t_{1}',t_{2}')$. 
		}
		\label{fig:mypath}
	\end{figure}
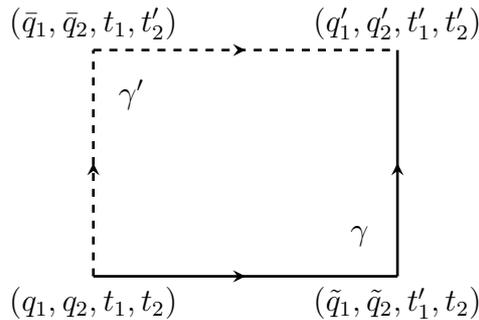

	To capture the quantum version of the consistency condition in terms of the Lagrangian, the appropriate approach is the Feynman path integration method. Let us first briefly provide some basic ingredients.
	\\
	\\
	\textbf{Single-time case}: The main mathematical object in this section is the propagator which is given by
	\begin{align}
		K(q_{f},t_{f};q_{i},t_{i})= \left\langle q_{f}\left| U(t_{f}-t_{i})\right|q_{i}\right\rangle\;.
	\end{align}
	The propagator provides the probability amplitude for a particle to travel from the initial point $(q_{i},t_{i})$ to the final point $(q_{f},t_{f})$. If we introduce the time $t_{1}$ such that $t_{f}>t_{1}>t_{i}$, the propagator can be factorised as follows
	\begin{align}
		K(q_{f},t_{f};q_{i},t_{i})&=\Big\langle q_{f}\Big| U(t_{f}-t_{1}+t_{1}-t_{i})\Big|q_{i}\Big\rangle \nonumber \\
		&= \Big\langle q_{f}\Big| U(t_{f}-t_{1})\int dq_{1}\Big| q_{1}\Big\rangle \Big\langle q_{1}\Big| U(t_{1}-t_{i})\Big|q_{i}\Big\rangle \nonumber \\
		&= \int dq_{1}K(q_{f},t_{f};q_{1},t_{1})K(q_{1},t_{1};q_{i},t_{i})\;.\label{K1}
	\end{align}
	Equation \eqref{K1} suggests that the transition amplitude of the quantum particle from the initial point to the final point must be taken into account of all possible points $q_1$ at time $t_1$. We could continue to make the time interval into $n$ parts such that $t_{n}\equiv t_{f}>t_{n-1}>t_{n-2}>...>t_{2}>t_{1}>t_{i}\equiv t_{0}$, resulting in

	\begin{equation}
		K(q_{n},t_{n};q_{0},t_{0})=\Bigg(\prod_{k=1}^{n-1}\int dq_{k}\Bigg) \prod_{k=0}^{n-1}K(q_{k+1},t_{k+1};q_{k},t_{k})\;.\label{propp}
	\end{equation}
	The discrete propagator is given by \cite{Feynman}
	\begin{align}
		K(q_{k+1},t_{k+1};q_{k},t_{k})
		&=\sqrt{\frac{m}{2\pi i(t_{k+1}-t_{k})}}e^{i(t_{k+1}-t_{k})L(q_{k},q_{k+1})}
		\;, \label{Kofk+1}
	\end{align}
	where
	\[L(q_{k},q_{k+1})=\frac{m}{2}\left(\frac{q_{k+1}-q_{k}}{t_{k+1}-t_{k}}\right)^{2}-V(q_{k}) \;.  \label{Ldiscon}\]
	Taking $t_{k+1}-t_{k} \equiv \Delta t \to 0$ and $n \to \infty $, 
	the propagator \eqref{propp} can be written as
	\begin{equation}
		K(q_{f},t_{f};q_{i},t_{i})=\int_{q_{i}}^{q_{f}} \mathscrsfs{D}[q(t)] e^{iS[q(t)]}  \;,  \label{eq:kk}
	\end{equation}
	where 
	\[\int_{q_{i}}^{q_{f}} \mathscrsfs{D}[q(t)]\equiv\lim_{\substack{n \to \infty \\ \Delta t \to 0}} \left(\sqrt{\frac{m}{2\pi i\Delta t}}\right)^{n/2}\left(\prod_{k=1}^{n-1}\int dq_{k}\right)\;,
	\]
	and 
	\[
	S[q(t)]=\int_{t_i}^{t_f}dtL(\dot q, q;t)\;.
	\]
	Here $L(\dot q, q;t)=T(\dot q)-V(q)$ is the standard single-time Lagrangian.
	\\
	\\
	\emph{Remark 3}: The explicit form of the propagator can be obtained for the case of quadratic Lagrangian
	\begin{equation}
		K\left(q_{f},t_{f};q_{i},t_{i}\right)=F(t_{f}-t_{i})e^{iS_{c}} \;, \label{eq:14}
	\end{equation}
	where $S_c$ is the classical action and $F(t_{f}-t_{i}) = \sqrt{\frac{1}{2\pi i}\Big|\frac{\partial^{2}S_{c}}{\partial{q_{i}}\partial{q_{f}'}}\Big|}$ is the prefactor \cite{Walter}.
	\\
	\\
	\textbf{Multi-time case}: Next, we would like to discuss the consistency condition for the multi-time evolution in the Feynman picture.
	\\
	\\
	\textbf{\emph{Compatible evolution}}: For simplicity, we consider the evolution of the multi-time wave function from the initial point $(t_{1},t_{2})$ to the final point $(t_{1}',t_{2}')$, see figure \ref{fig:mypath}, from two different paths in the context of Feynman path integration on the space of time variables.
	\\
	\\
	The first path(solid line): The transition of the multi-time wave function from point $(t_{1},t_{2})$ to $(t_{1}',t_{2}')$ evolves from $t_1$ to $t'_1$ with the unitary operator $U_1$, then evolves from $t_2$ to $t'_2$ with the unitary operator $U_2$. The lower-half path can be captured in terms of the propagator as follows
	%
	%
	\begin{align}
		\left\langle q'_{1},q'_{2}\big|\Phi_{\lrcorner}(t'_{1},t'_{2})\right\rangle&=\left\langle q'_{1},q'_{2}\left|U_{2}U_{1}\right|\Phi(t_{1},t_{2})\right\rangle \nonumber \\
		&=\iiiint d\tilde{q}_{1} d\tilde{q}_{2} dq_{1} dq_{2} \left\langle q'_{1},q'_{2}\left|U_{2}\right|\tilde{q}_{2},\tilde{q}_{1}\right\rangle \left\langle\tilde{q}_{2},\tilde{q}_{1}\left|U_{1}\right|q_{1},q_{2}\right\rangle\left\langle q_{2},q_{1}|\Phi(t_{1},t_{2})\right\rangle \nonumber \\
		&=\iiiint d\tilde{q}_{1} d\tilde{q}_{2} dq_{1} dq_{2} \left\langle q'_{2}\left|U_{2}\right|\tilde{q}_{2}\right\rangle \left\langle q'_{1}|\tilde{q}_{1}\right\rangle \left\langle \tilde{q}_{1}\left|U_{1}\right|q_{1}\right\rangle \left\langle \tilde{q}_{2}|q_{2}\right\rangle \Phi(q_{1},q_{2},t_{1},t_{2}) \nonumber \\
		&=\iiiint d\tilde{q}_{1} d\tilde{q}_{2} dq_{1} dq_{2} \left\langle q'_{2}\left|U_{2}\right|\tilde{q}_{2}\right\rangle \left\langle \tilde{q}_{1}\left|U_{1}\right|q_{1}\right\rangle \delta(q'_{1}-\tilde{q}_{1}) \delta(\tilde{q}_{2}-q_{2}) \Phi(q_{1},q_{2},t_{1},t_{2}) \nonumber \\
		&=\iint dq_{1} dq_{2} \left\langle q'_{2}\left|U_{2}\right|q_{2}\right\rangle \left\langle q'_{1}\left|U_{1}\right|q_{1}\right\rangle  \Phi(q_{1},q_{2},t_{1},t_{2}) \nonumber \\
		\Phi_{\lrcorner} \left(q_{1}',q_{2}',t_{1}',t_{2}'\right)&=\iint dq_{1} dq_{2} K_{2}(q_{2}',t_{2}';q_{2},t_{2}) K_{1}(q_{1}',t_{1}';q_{1},t_{1}) \Phi(q_{1},q_{2},t_{1},t_{2})
		\;. \label{pathindependent1}
	\end{align}
	The second path(dashed line): The transition of the multi-time wave function from point $(t_{1},t_{2})$ to $(t_{1}',t_{2}')$ through the upper-half path is given by
	\begin{align}
		\left\langle q'_{1},q'_{2}\big|\Phi_{\ulcorner}(t'_{1},t'_{2})\right\rangle&=\left\langle q'_{1},q'_{2}\left|U_{1}U_{2}\right|\Phi(t_{1},t_{2})\right\rangle \nonumber \\
		&=\iiiint d\bar{q}_{1} d\bar{q}_{2} dq_{1} dq_{2} \left\langle q'_{1},q'_{2}\left|U_{1}\right|\bar{q}_{2},\bar{q}_{1}\right\rangle \left\langle\bar{q}_{2},\bar{q}_{1}\left|U_{2}\right|q_{1},q_{2}\right\rangle\left\langle q_{2},q_{1}|\Phi(t_{1},t_{2})\right\rangle \nonumber \\
		&=\iiiint d\bar{q}_{1} d\bar{q}_{2} dq_{1} dq_{2} \left\langle q'_{1}\left|U_{1}\right|\bar{q}_{1}\right\rangle \left\langle q'_{2}|\bar{q}_{2}\right\rangle \left\langle \bar{q}_{2}\left|U_{2}\right|q_{2}\right\rangle \left\langle \bar{q}_{1}|q_{1}\right\rangle \Phi(q_{1},q_{2},t_{1},t_{2}) \nonumber \\
		&=\iiiint d\bar{q}_{1} d\bar{q}_{2} dq_{1} dq_{2} \left\langle q'_{1}\left|U_{1}\right|\bar{q}_{1}\right\rangle \left\langle \bar{q}_{2}\left|U_{2}\right|q_{2}\right\rangle \delta(q'_{2}-\bar{q}_{2}) \delta(\bar{q}_{1}-q_{1}) \Phi(q_{1},q_{2},t_{1},t_{2}) \nonumber \\
		&=\iint dq_{1} dq_{2} \left\langle q'_{1}\left|U_{1}\right|q_{1}\right\rangle \left\langle q'_{2}\left|U_{2}\right|q_{2}\right\rangle \Phi(q_{1},q_{2},t_{1},t_{2}) \nonumber \\
		\Phi_{\ulcorner} \left(q_{1}',q_{2}',t_{1}',t_{2}'\right)&=\iint dq_{1} dq_{2} K_{1}(q_{1}',t_{1}';q_{1},t_{1}) K_{2}(q_{2}',t_{2}';q_{2},t_{2}) \Phi(q_{1},q_{2},t_{1},t_{2})
		\;. \label{pathindependent2}
	\end{align}
	%
	To make the both transitions compatible, one requires $\Phi_{\lrcorner} \left(q_{1}',q_{2}',t_{1}',t_{2}'\right)=\Phi_{\ulcorner} \left(q_{1}',q_{2}',t_{1}',t_{2}'\right)$, resulting in
	\begin{align}
		\iint dq_{1} dq_{2} \left\{K_{2}K_{1}-K_{1}K_{2}\right\} \Phi(q_{1},q_{2},t_{1},t_{2}) &= 0 \;.\label{eq:47}
	\end{align}
	If now we define $K_{\lrcorner}\left(q'_{1},t'_1,q'_{2},t'_2;q_{1},t_{1},q_{2},t_{2}\right)=K_{2}(q_{2}',t_{2}';q_{2},t_{2}) K_{1}(q_{1}',t_{1}';q_{1},t_{1})$ as a lower-half propagator and $K_{\ulcorner}\left(q'_{1},t'_1,q'_{2},t'_2;q_{1},t_{1},q_{2},t_{2}\right)=K_{1}(q_{1}',t_{1}';q_{1},t_{1}) K_{2}(q_{2}',t_{2}';q_{2},t_{2})$ as an upper-half propagator. Since $\Phi(q_{1},q_{2},t_{1},t_{2})$ cannot be zero, then \eqref{eq:47} gives us
	\begin{align}
		K_{\lrcorner}\left(q'_{1},t'_1,q'_{2},t'_2;q_{1},t_{1},q_{2},t_{2}\right)=K_{\ulcorner}\left(q'_{1},t'_1,q'_{2},t'_2;q_{1},t_{1},q_{2},t_{2}\right)
		\;. \label{pathindependent3}
	\end{align}
	Here we obtain the consistency condition for the multi-time evolution in terms of the propagator. This equation is nothing but the commuting propagators: $[K_1,K_2 ]=0$ reflecting the path-independent property of the propagator on the space of time variables.
	\\\\
	One can treat these commuting propagators \eqref{pathindependent3} as the parallel transport operation in terms of Lagrangian. Now we may write the propagator in terms of the Wilson line associated with path $\gamma$ as
	\begin{align}
		K_{\gamma}\left(q'_{1},t'_1,q'_{2},t'_2;q_{1},t_{1},q_{2},t_{2}\right)=\int_{q_{2}}^{q'_{2}} \mathscrsfs{D}[\tilde{q}_2(\tilde{t}_2)]\int_{q_{1}}^{q'_{1}} \mathscrsfs{D}[\tilde{q}_1(\tilde{t}_1)]e^{i\int_{\gamma}L_1(\tilde{q}_1,\partial_{\tilde{t}_1}\tilde{q_1})d\tilde{t}_1+L_2(\tilde{q}_2,\partial_{\tilde{t}_2}\tilde{q_2})d\tilde{t}_2}
		\; \label{pathindependent322}
	\end{align}
	and the Wilson line associated with path $\gamma'$ as
	\begin{align}
		K_{\gamma'}\left(q'_{1},t'_1,q'_{2},t'_2;q_{1},t_{1},q_{2},t_{2}\right)=\int_{q_{1}}^{q'_{1}} \mathscrsfs{D}[\tilde{q}_1(\tilde{t}_1)]\int_{q_{2}}^{q'_{2}} \mathscrsfs{D}[\tilde{q}_2(\tilde{t}_2)]e^{i\int_{\gamma'}L_1(\tilde{q}_1,\partial_{\tilde{t}_1}\tilde{q_1})d\tilde{t}_1+L_2(\tilde{q}_2,\partial_{\tilde{t}_2}\tilde{q_2})d\tilde{t}_2}
		\;. \label{pathindependent32a2}
	\end{align}
	Equation \eqref{pathindependent3} gives invariant property of the propagator sharing the end-points $K_{\gamma}=K_{\gamma'}$. The result in \eqref{pathindependent322} and \eqref{pathindependent32a2} can be easily extended to the case of $N$ time variables and the Wilson line $\gamma$ in terms of the propagator is given by
	\begin{align}
		K_{\gamma}\left(q'_{1},t'_1,q'_{2},t'_2,...,q'_{N},t'_N;q_{1},t_{1},q_{2},t_{2},...,q_{N},t_{N}\right)=\mathsf{P}\prod_{i=1}^N\int_{q_{i}}^{q'_{i}} \mathscrsfs{D}[\tilde{q}_i(\tilde{t}_i)]e^{i\int_{\gamma}\sum_{i=1}^NL_i(\tilde{q}_i,\partial_{\tilde{t}_i}\tilde{q_i})d\tilde{t}_i}
		\;, \label{pathindependent3ss22}
	\end{align}
	where $\mathsf P$ stands for the permutation. Therefore, the propagator \eqref{pathindependent3ss22} is invariant under the permutation.
	\\
	\\
	\textbf{\emph{Time loops}}: 
	We now consider another type of evolution called the loop transition. Before proceeding the calculation, we need to establish some useful relations. We start to consider the transition of the wave function from $(q,t)$ to $(q',t')$ given by
	\begin{align}
		\Phi(q',t')&=\int dq K(q',t';q,t) \Phi(q,t)
		\;. \label{koffree3}
	\end{align}
	Next, we consider the transition from $(q',t')$ to $(\tilde{q},\tilde{t})$ given by
	\begin{align}
		\Phi(\tilde{q},\tilde{t})&=\int dq' K(\tilde{q},\tilde{t};q',t') \Phi(q',t')
		\;. \label{koffree6}
	\end{align}
	Combining (\ref{koffree6}) with (\ref{koffree3}), we obtain
	\begin{align}
		\Phi(\tilde{q},\tilde{t})&=\iint dq' dq K(\tilde{q},\tilde{t};q',t') K(q',t';q,t) \Phi(q,t)
		\;. \label{koffree7}
	\end{align}
	To change the transition \eqref{koffree7} to the loop transition, we impose
	\begin{align}
		\Phi(\tilde{q},\tilde{t})&=\iint dq' dq K(\tilde{q},\tilde{t};q',t') K(q',t';q,t) \Phi(q,t) =\int dq \delta(\tilde{q}-q) \Phi(q,\tilde{t}) =\Phi(\tilde{q},\tilde{t})
		\;, \label{koffree8new}
	\end{align}
	therefore, one requires
	\begin{align}
		\delta(\tilde{q}-q)&=\int dq' K(\tilde{q},\tilde{t};q',t') K(q',t';q,t) =K(\tilde{q},t;q,t)
		\;, \label{koffree9new}
	\end{align}
	where $(\tilde{t}-t)= \delta t \to 0$. Equivalently, \eqref{koffree9new} can be expressed in terms of Lagrangians as
	\begin{align}
		\delta(\tilde{q}-q)&=\lim_{\delta t \to 0} \int dq' \left[\int_{q'}^{\tilde{q}}\mathscrsfs{D}[\bar{q}(\bar{t})] e^{i\int_{t'}^{\tilde{t}}L(\bar{q},\partial_{\bar{t}}\bar{q})d\bar{t}}\right]\left[\int_{q}^{q'}\mathscrsfs{D}[\bar{q}(\bar{t})] e^{i\int_{t}^{t'}L(\bar{q},\partial_{\bar{t}}\bar{q})d\bar{t}}\right] \nonumber  \\
		&=\lim_{\delta t \to 0} \int dq' \int_{q'}^{\tilde{q}}\mathscrsfs{D}[\bar{q}(\bar{t})] \int_{q}^{q'}\mathscrsfs{D}[\bar{q}(\bar{t})] e^{i\left(\int_{t'}^{\tilde{t}}L(\bar{q},\partial_{\bar{t}}\bar{q})d\bar{t}+\int_{t}^{t'}L(\bar{q},\partial_{\bar{t}}\bar{q})d\bar{t}\right)} \nonumber  \\
		&=\lim_{\delta t \to 0} \int_{q}^{\tilde{q}}\mathscrsfs{D}[\bar{q}(\bar{t})]e^{i\int_{t}^{\tilde{t}} L(\bar{q},\partial_{\bar{t}}\bar{q})d\bar{t}}=\int_{q}^{\tilde{q}}\mathscrsfs{D}[\bar{q}(\bar{t})]e^{i\oint L(\bar{q},\partial_{\bar{t}}\bar{q})d\bar{t}}
		\;. \label{koffree10new2}
	\end{align}
	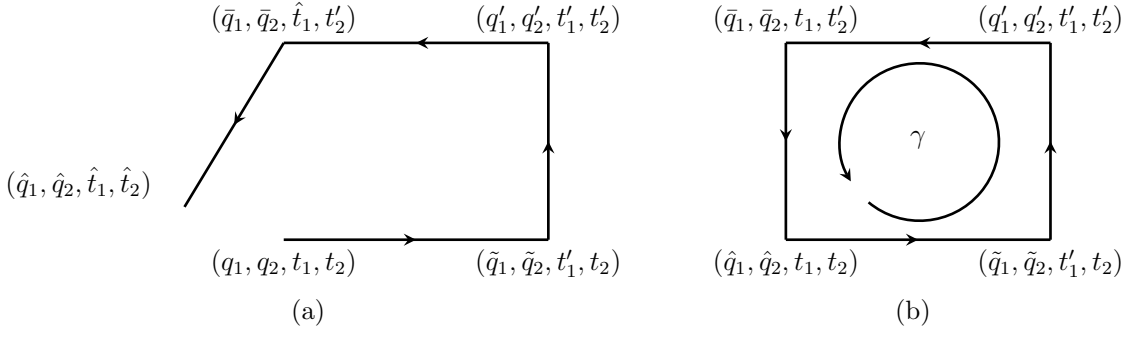
\begin{figure}[h]
		\centering
		\subfloat[]{\label{fig:loop2timeA}
			\begin{tikzpicture}[scale=0.87, every node/.style={transform shape}]
				\foreach \coord/\label [count=\xi] in {
					{0,0}/{$(q_{1},q_{2},t_{1},t_{2})$},
					{4,3}/{$(q'_{1},q'_{2},t'_{1},t'_{2})$},
					{4,0}/{$(\tilde{q}_{1},\tilde{q}_{2},t'_{1},t_{2})$},
					{0,3}/{$(\bar{q}_{1},\bar{q}_{2},\hat{t}_{1},t'_{2})$},
					{-3.1,1.25}/{$(\hat{q}_{1},\hat{q}_{2},\hat{t}_{1},\hat{t}_{2})$}
				}{
					\pgfmathsetmacro\anch{mod(\xi,2) ? "north" : "south"}
					\node[anchor=\anch] at (\coord) {\label};
				}
				\draw[line width=1pt,postaction={on each segment={mid arrow}} ] (0,0) -- (4,0) ;
				\draw[line width=1pt,postaction={on each segment={mid arrow}} ] (4,0) -- (4,3) ;
				\draw[line width=1pt,postaction={on each segment={mid arrow}} ] (4,3) -- (0,3) ;
				\draw[line width=1pt,postaction={on each segment={mid arrow}} ] (0,3) -- (-1.5,0.5) ;
			\end{tikzpicture}%
		}
		\qquad
		\subfloat[]{\label{fig:loop2timeB}
			\begin{tikzpicture}[scale=0.87, every node/.style={transform shape}]
				\foreach \coord/\label [count=\xi] in {
					{0,0}/{$(\hat{q}_{1},\hat{q}_{2},t_{1},t_{2})$},
					{4,3}/{$(q'_{1},q'_{2},t'_{1},t'_{2})$},
					{4,0}/{$(\tilde{q}_{1},\tilde{q}_{2},t'_{1},t_{2})$},
					{0,3}/{$(\bar{q}_{1},\bar{q}_{2},t_{1},t'_{2})$},
					{2,1.8}/{$\gamma$}
				}{
					\pgfmathsetmacro\anch{mod(\xi,2) ? "north" : "south"}
					\node[anchor=\anch] at (\coord) {\label};
				}
				\draw[line width=1pt,postaction={on each segment={mid arrow}} ] (0,0) -- (4,0);
				\draw[line width=1pt,postaction={on each segment={mid arrow}} ] (4,0) -- (4,3);
				\draw[line width=1pt,postaction={on each segment={mid arrow}} ] (4,3) -- (0,3);
				\draw[line width=1pt,postaction={on each segment={mid arrow}} ] (0,3) -- (0,0);
				\draw[-stealth,line width=1pt] (1.25,0.57) arc (50:390:-1.2);
			\end{tikzpicture}%
		}
		\caption{(a) The evolution from the initial point $(q_1,q_2,t_1,t_2)$ to the final point $(\hat q_1,\hat q_2,\hat t_1,\hat t_2)$. (b) A loop evolution $\gamma$ can be obtained by imposing $\hat q_i=q_i$ and $\hat t_i=t_i$, where $i=1,2$.}
		\label{fig:loop2time}
	\end{figure}
	\\
	Now we are ready to consider the loop evolution. Let's define $U_1(t_1'-t_1)$ as the time evolution operator from $t_1$ to $t'_1$, $U_2(t_2'-t_2)$ as the time evolution operator from $t_2$ to $t'_2$, $U'_1(\hat t_1-t'_1)$ as the time evolution operator from $t'_1$ to $\hat t_1$ and $U_2'(\hat t_2-t_2')$ as the time evolution operator from $t'_2$ to $\hat t_2$.
	The transition map, shown in figure \ref{fig:loop2timeA}, can be expressed as
	\begin{align}
		\left\langle \hat{q}_{1},\hat{q}_{2}\big|\Phi(\hat{t}_{1},\hat{t}_{2})\right\rangle &= \left\langle \hat{q}_{1},\hat{q}_{2}\left|U_{2}'U_{1}'U_{2}U_{1}\right|\Phi(t_{1},t_{2})\right\rangle \nonumber  \\
		\Phi\left(\hat{q}_{1},\hat{q}_{2},\hat{t}_{1},\hat{t}_{2}\right)&=\iint dq_{1} dq_{2} \left\langle \hat{q}_{1},\hat{q}_{2}\left|U_{2}'U_{1}'U_{2}U_{1}\right|q_{1},q_{2}\right\rangle\left\langle q_{2},q_{1}\big|\Phi(t_{1},t_{2})\right\rangle \nonumber  \\
		&=\iint dq_{1} dq_{2} \int d\bar{q}_{2} \left\langle \hat{q}_{2}\left|U_{2}'(\hat{t}_{2}-t'_{2})\right|\bar{q}_{2}\right\rangle  \left\langle \bar{q}_{2}\left|U_{2}(t'_{2}-t_{2})\right|q_{2}\right\rangle \nonumber  \\
		&\quad\cross \int d\tilde{q}_{1} \left\langle \hat{q}_{1} \left|U_{1}'(\hat{t}_{1}-t'_{1})\right|\tilde{q}_{1}\right\rangle  \left\langle \tilde{q}_{1}\left|U_{1}(t'_{1}-t_{1})\right|q_{1}\right\rangle \Phi(q_{1},q_{2},t_{1},t_{2})  \nonumber  \\
		&=\iint dq_{1} dq_{2} \int d\bar{q}_{2} K_{2}(\hat{q}_{2},\hat{t}_{2};\bar{q}_{2},t'_{2}) K_{2}(\bar{q}_{2},t'_{2};q_{2},t_{2}) \int d\tilde{q}_{1} K_{1}(\hat{q}_{1},\hat{t}_{1};\tilde{q}_{1},t'_{1}) K_{1}(\tilde{q}_{1},t'_{1};q_{1},t_{1}) \nonumber  \\
		&\quad\cross \Phi(q_{1},q_{2},t_{1},t_{2})
		\;. \label{loop2timenewnew}
	\end{align}
	The full derivation of \eqref{loop2timenewnew} can be found in the appendix. Using the condition \eqref{koffree9new} where $(\hat{t}_1-t_1)=\delta t_1 \to 0$ and $(\hat{t}_2-t_2)=\delta t_2 \to 0$, we have
	\begin{align}
		\delta(\hat{q}_{2}-q_{2})&=\int d\bar{q}_{2} K_{2}(\hat{q}_{2},\hat{t}_{2};\bar{q}_{2},t'_{2}) K_{2}(\bar{q}_{2},t'_{2};q_{2},t_{2}) = K_2(\hat{q}_{2},t_{2};q_{2},t_{2}) \;, \label{loop2time21new} \\
		\delta(\hat{q}_{1}-q_{1})&=\int d\tilde{q}_{1} K_{1}(\hat{q}_{1},\hat{t}_{1};\tilde{q}_{1},t'_{1}) K_{1}(\tilde{q}_{1},t'_{1};q_{1},t_{1}) = K_1(\hat{q}_{1},t_{1};q_{1},t_{1})
		\;. \label{loop2time2new}
	\end{align}
	Substituting \eqref{loop2time21new} and \eqref{loop2time2new} into \eqref{loop2timenewnew}, we find that
	\begin{align}
		\Phi\left(\hat{q}_{1},\hat{q}_{2},t_{1},t_{2}\right)&=\iint dq_{1} dq_{2} \delta(\hat{q}_{2}-q_{2}) \delta(\hat{q}_{1}-q_{1}) \Phi(q_{1},q_{2},t_{1},t_{2})=\Phi(\hat{q}_{1},\hat{q}_{2},t_{1},t_{2})
		\;, \label{loopconsistency}
	\end{align}
	which gives us the loop evolution shown in figure \ref{fig:loop2timeB}.
	\\
	\\
	Next, the condition for the propagator in \eqref{loop2timenewnew} can be expressed in terms of the Lagrangian as
	\begin{align}
		\delta\left(\hat{q}_{2}-q_{2}\right)  \delta\left(\hat{q}_{1}-q_{1}\right)&= \int d\bar{q}_{2} K_{2}(\hat{q}_{2},\hat{t}_{2};\bar{q}_{2},t'_{2}) K_{2}(\bar{q}_{2},t'_{2};q_{2},t_{2}) \int d\tilde{q}_{1} K_{1}(\hat{q}_{1},\hat{t}_{1};\tilde{q}_{1},t'_{1}) K_{1}(\tilde{q}_{1},t'_{1};q_{1},t_{1}) \nonumber  \\
		&= \int d\bar{q}_{2} \left[\int_{\bar{q_{2}}}^{\hat{q}_{2}}\mathscrsfs{D}[\check{q_{2}}(\check{t}_{2})] e^{i\int_{t'_{2}}^{\hat{t}_{2}}L_{2}(\check{q_{2}},\partial_{\check{t}_2}\check{q}_2)d\check{t}_{2}}\right] \left[\int_{q_{2}}^{\bar{q_{2}}}\mathscrsfs{D}[\check{q_{2}}(\check{t}_{2})] e^{i\int_{t_{2}}^{t'_{2}}L_{2}(\check{q_{2}},\partial_{\check{t}_2}\check{q}_2)d\check{t}_{2}}\right] \nonumber  \\
		&\quad\cross \int d\tilde{q}_{1} \left[\int_{\tilde{q_{1}}}^{\hat{q}_{1}}\mathscrsfs{D}[\check{q_{1}}(\check{t}_{1})] e^{i\int_{t'_{1}}^{\hat{t}_{1}}L_{1}(\check{q_{1}},\partial_{\check{t}_1}\check{q}_1)d\check{t}_{1}}\right] \left[\int_{q_{1}}^{\tilde{q_{1}}}\mathscrsfs{D}[\check{q_{1}}(\check{t}_{1})] e^{i\int_{t_{1}}^{t'_{1}}L_{1}(\check{q_{1}},\partial_{\check{t}_1}\check{q}_1)d\check{t}_{1}}\right] \nonumber  \\
		&= \left[  \int_{q_{2}}^{\hat{q}_{2}}\mathscrsfs{D}[\check{q_{2}}(\check{t}_{2})] e^{i\int_{t_{2}}^{\hat{t}_{2}} L_{2}(\check{q_{2}},\partial_{\check{t}_2}\check{q}_2)d\check{t}_{2}}\right] \left[  \int_{q_{1}}^{\hat{q}_{1}}\mathscrsfs{D}[\check{q_{1}}(\check{t}_{1})] e^{i\int_{t_{1}}^{\hat{t}_{1}} L_{1}(\check{q_{1}},\partial_{\check{t}_1}\check{q}_1)d\check{t}_{1}}\right]
		\;. \label{loopk}
	\end{align}
	Taking $\delta t_1 \to 0$ and $\delta t_2 \to 0$, we obtain
	\begin{align}
		\delta\left(\hat{q}_{2}-q_{2}\right)  \delta\left(\hat{q}_{1}-q_{1}\right)
		&= \left[ \lim_{\delta t_2 \to 0} \int_{q_{2}}^{\hat{q}_{2}}\mathscrsfs{D}[\check{q_{2}}] e^{i\int_{t_{2}}^{\hat{t}_{2}} L_{2}(\check{q_{2}},\partial_{\check{t}_2}\check{q}_2)d\check{t}_{2}}\right] \left[ \lim_{\delta t_1 \to 0}  \int_{q_{1}}^{\hat{q}_{1}}\mathscrsfs{D}[\check{q_{1}}] e^{i\int_{t_{1}}^{\hat{t}_{1}} L_{1}(\check{q_{1}},\partial_{\check{t}_1}\check{q}_1)d\check{t}_{1}}\right] \nonumber  \\
		&= \left[  \int_{q_{2}}^{\hat{q}_{2}}\mathscrsfs{D}[\check{q_{2}}] e^{i\oint L_{2}(\check{q_{2}},\partial_{\check{t}_2}\check{q}_2)d\check{t}_{2}}\right] \left[ \int_{q_{1}}^{\hat{q}_{1}}\mathscrsfs{D}[\check{q_{1}}] e^{i\oint L_{1}(\check{q_{1}},\partial_{\check{t}_1}\check{q}_1)d\check{t}_{1}}\right]
		\;. \label{loopk}
	\end{align}
	The expression in \eqref{loopk} can be immediately extended to the case of $N$ time variables resulting in
	\begin{align}
		\prod_{k=1}^{N} \int_{q_{k}}^{\hat{q}_{k}} \mathscrsfs{D}[\check{q_{k}}] e^{i\oint_{\gamma} L_{k}(\check{q_{k}},\partial_{\check{t}_k}\check{q}_k)d\check{t}_{k}} = \prod_{k=1}^{N} \delta (\hat{q}_{k}-q_{k})
		\;. \label{loopk2}
	\end{align}
	In the language of the Wilson line, we have the propagator for the loop $\gamma$ as
	\begin{align}
		K_{\gamma}(\hat{q}_1,t_1,\hat{q}_2,t_2,...,\hat{q}_N,t_N;\hat{q}_1,t_1,\hat{q}_2,t_2,...,\hat{q}_N,t_N)=\prod_{k=1}^{N} \oint \mathscrsfs{D}[\check{q_{k}}] e^{i\oint_{\gamma} L_{k}(\check{q_{k}},\partial_{\check{t}_k}\check{q}_k)d\check{t}_{k}} = I
		\;. \label{loopk23}
	\end{align}
	\\
	What we have from \eqref{loopk23} is the following. The quantum transition between two endpoints will get no contribution from the loops. In other words, the loops can be excluded from the whole evolution as shown in figure \ref{fig:loopindependent}.
	\begin{figure}[h]
		\centering
		\begin{tikzpicture}[scale=0.6]
			\foreach \coord/\label [count=\xi] in {
				{5.4,1.35}/{$t_{1}$},
				{-1,6}/{$t_{2}$}	
			}
			{
				\pgfmathsetmacro\anch{mod(\xi,2) ? "north" : "south"}
				\node[anchor=\anch] at (\coord) {\label};
			}
			
			\draw[-stealth,line width=1.25pt] (-1,1) -- (5,1);
			\draw[-stealth,line width=1.25pt] (-1,1) -- (-1,6);
			
			\draw[-stealth,line width=1pt] (0,2) -- (0,3);
			\draw[-stealth,line width=1pt] (0,3) -- (2,3);
			\draw[-stealth,line width=1pt] (2,3) -- (2,5);
			\draw[-stealth,line width=1pt] (2,5) -- (1,5);
			\draw[-stealth,line width=1pt] (1,5) -- (1,4);
			\draw[-stealth,line width=1pt] (1,4) -- (3.5,4);
			
		\end{tikzpicture}
		\qquad
		\begin{tikzpicture}[scale=0.6]
			\foreach \coord/\label [count=\xi] in {
				{5.4,1.35}/{$t_{1}$},
				{-1,6}/{$t_{2}$}	
			}
			{
				\pgfmathsetmacro\anch{mod(\xi,2) ? "north" : "south"}
				\node[anchor=\anch] at (\coord) {\label};
			}
			
			\draw[-stealth,line width=1.25pt] (-1,1) -- (5,1);
			\draw[-stealth,line width=1.25pt] (-1,1) -- (-1,6);
			
			\draw[-stealth,line width=1pt] (0,2) -- (0,3);
			\draw[-stealth,line width=1pt] (0,3) -- (2,3);
			\draw[-stealth,line width=1pt] (2,3) -- (2,4);
			\draw[-stealth,line width=1pt] (2,4) -- (3.5,4);
			
		\end{tikzpicture}
		\caption{The close loop does not contribute to the evolution.}
		\label{fig:loopindependent}
	\end{figure}
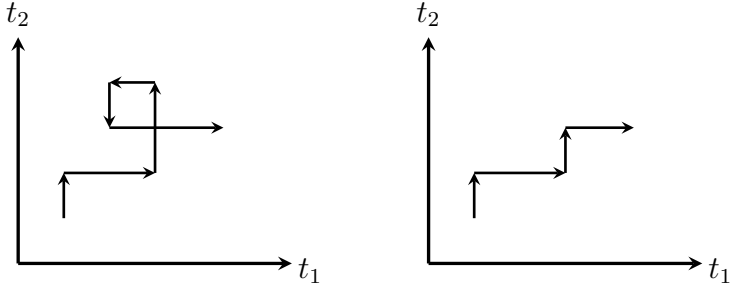
	\\
	\emph{Example}: Next, we will give an explicit computation to illustrate the path independent property, i.e., a loop evolution. Here, for simplicity, we choose a free particle to work with and the propagator is given by
	\begin{align}
		K(q',t';q,t)&=\sqrt{\frac{m}{2\pi i  (t'-t)}}e^{\frac{im}{2}\frac{(q'-q)^2}{(t'-t)}} \;.  
		\label{freepropagator}
	\end{align}
	%
	We now compute the propagator along the time variables $t_i$, where $i=1,2$
	\begin{align}
		K_i(\hat{q}_i,\hat{t}_i;q_i,t_i)&=\int d\bar{q}_i K_i(\hat{q}_i,\hat{t}_i;\bar{q}_i,t'_i) K_i(\bar{q}_i,t'_i;q_i,t_i) \nonumber  \\
		&=\int d\bar{q}_i \sqrt{\frac{m}{2i\pi(\hat{t}_i-t'_i)}} \sqrt{\frac{m}{2i\pi(t'_i-t_i)}} e^{\frac{im}{2}\frac{(\hat{q}_i-\bar{q}_i)^2}{(\hat{t}_i-t'_i)}} e^{\frac{im}{2}\frac{(\bar{q}_i-q_i)^2}{(t'_i-t_i)}} \nonumber  \\
		&=\int d\bar{q}_i \frac{m}{2i\pi} \sqrt{\frac{1}{(\hat{t}_i-t'_i)(t'_i-t_i)}} e^{\frac{im}{2(\hat{t}_i-t'_i)}(\hat{q}_i^2-2\hat{q}_i\bar{q}_i+\bar{q}_i^2)} e^{\frac{im}{2(t'_i-t_i)}(\bar{q}_i^2-2\bar{q}_iq_i+q_i^2)} \nonumber  \\
		&=\int d\bar{q}_i \frac{m}{2i\pi} \sqrt{\frac{1}{(\hat{t}_i-t'_i)(t'_i-t_i)}} e^{\hat{q}_i^2\left(\frac{im}{2(\hat{t}_i-t'_i)}\right)} e^{q_i^2\left(\frac{im}{2(t'_i-t_i)}\right)} e^{\bar{q}_i^2\left(\frac{im}{2(\hat{t}_i-t'_i)}+\frac{im}{2(t'_i-t_i)}\right)} e^{\bar{q}_i\left(\frac{-im\hat{q}_i}{(\hat{t}_i-t'_i)}+\frac{-imq_i}{(t'_i-t_i)}\right)} \nonumber  \\
		&=\frac{m}{2i\pi} \sqrt{\frac{2\pi}{(-im)(\hat{t}_i-t_i)}} e^{\hat{q}_i^2\left(\frac{im}{2(\hat{t}_i-t'_i)}\right)} e^{q_i^2\left(\frac{im}{2(t'_i-t_i)}\right)} e^{\frac{(-im)^2}{-2im}\left(\frac{\hat{q}_i}{(\hat{t}_i-t'_i)}+\frac{q_i}{(t'_i-t_i)}\right)^2\left(\frac{(\hat{t}_i-t'_i)(t'_i-t_i)}{(\hat{t}_i-t_i)}\right)} \nonumber  \\
		&=\sqrt{\frac{m}{2i\pi(\hat{t}_i-t_i)}} 
		e^{\hat{q}_i^2 \frac{im}{2(\hat{t}_i-t'_i)}\left(1-\frac{(t'_i-t_i)}{(\hat{t}_i-t_i)}\right)} 
		e^{q_i^2 \frac{im}{2(t'_i-t_i)}\left(1-\frac{(\hat{t}_i-t'_i)}{(\hat{t}_i-t_i)}\right)} 
		e^{2\hat{q}_iq_i\left(\frac{-im}{2(\hat{t}_i-t_i)}\right)}  \nonumber  \\
		&=\sqrt{\frac{m}{2i\pi(\hat{t}_i-t_i)}} 
		e^{\frac{im}{2(\hat{t}_i-t_i)}\left(\hat{q}_i^2-2\hat{q}_iq_i+q_i^2\right)}   \nonumber  \\
		&=\sqrt{\frac{m}{2i\pi(\hat{t}_i-t_i)}} 
		e^{\frac{im}{2(\hat{t}_i-t_i)}\left(\hat{q}_i-q_i\right)^2}
		\;. \label{loopKtwo}
	\end{align}
	Imposing $\hat{t}_i-t_i=\delta t_i$ and taking $\delta t_i \to 0$, we obtain \cite{Walter}
	\begin{align}
		K_i(\hat{q}_i,t_i;q_i,t_i)&=\lim_{\delta t_i \to 0} \sqrt{\frac{m}{2i\pi\delta t_i}}
		e^{\frac{im}{2\delta t_i}\left(\hat{q}_i-q_i\right)^2}=\delta (\hat{q}_i-q_i)
		\;, \label{loopKtwo2}
	\end{align}
	which are indeed \eqref{loop2time21new} for $i=2$ and \eqref{loop2time2new} for $i=1$.
	%
	\\
	\\
	\textbf{\emph{Including interaction}}: The last point is that we will consider the system with the interaction. For simplicity, we work with the Hamiltonian for the system of two particles
	\begin{align}
		H=H_{1}+H_{2}+V_{12} \;,
	\end{align}
	where $V_{12}$ is a potential representing the interaction between the particles and $H_i$ is the free Hamiltonian for the $i^{\text{th}}$ particle. What we are going to do is the same process as in figure \ref{fig:mypath}.
	\\
	\\
	Let us first define the unitary operators $
	U_{1}(t'_{1},t_{1})=e^{-i{(H_{1}+V_{12})}(t'_1-t_{1})}$ and $
	U_{2}(t'_{2},t_{2})=e^{-i{H_{2}}(t'_2-t_{2})}$. Then the propagator for the lower corner path is given by
	\textcolor{black}{
	\begin{align}
		K_{\lrcorner}\left(q'_{1},t'_1,q'_{2},t'_2;q_{1},t_{1},q_{2},t_{2}\right)&=\left\langle q'_{1},q'_{2}\left|U_{2}U_{1}\right|q_{1},q_{2}\right\rangle \nonumber \\
		&=\iint d\tilde{q}_{1} d\tilde{q}_{2} \left\langle q'_{1},q'_{2}\left|U_{2}\right|\tilde{q}_{2},\tilde{q}_{1}\right\rangle \left\langle\tilde{q}_{2},\tilde{q}_{1}\left|U_{1}\right|q_{1},q_{2}\right\rangle \nonumber \\
		&=\iint d\tilde{q}_{1} d\tilde{q}_{2} \left\langle q'_{1}|\tilde{q}_{1}\right\rangle \left\langle q'_{2} \left|U_{2}\right|\tilde{q}_{2}\right\rangle \left\langle \tilde{q}_{2},\tilde{q}_{1} \left|U_{1}\right|q_{1},q_{2}\right\rangle \nonumber \\
		&=\iint d\tilde{q}_{1} d\tilde{q}_{2} \delta(q'_{1}-\tilde{q}_{1}) K_{2}(q'_{2},t'_{2};\tilde{q}_{2},t_{2}) \left\langle \tilde{q}_{2},\tilde{q}_{1} \left|U_{1}\right|q_{1},q_{2}\right\rangle \nonumber \\
		&=\int d\tilde{q}_{2} K_{2}(q'_{2},t'_{2};\tilde{q}_{2},t_{2}) \left\langle \tilde{q}_{2}\Big|\left\langle q'_{1} \left|U_{1}\right|q_{1}\right\rangle\Big|q_{2}\right\rangle \nonumber \\
		&=\int d\tilde{q}_{2} K_{2}(q'_{2},t'_{2};\tilde{q}_{2},t_{2}) \left\langle \tilde{q}_{2}\left| G(q'_{1},t'_{1};q_{1},t_{1};\hat{q_{2}})\right|q_{2}\right\rangle \nonumber \\
		&=\int d\tilde{q}_{2} K_{2}(q'_{2},t'_{2};\tilde{q}_{2},t_{2}) \left\langle \tilde{q}_{2}|q_{2}\right\rangle G(q'_{1},t'_{1};q_{1},t_{1};\tilde{q}_{2},t_{2};q_{2},t_{2})  \nonumber \\
		&=\int d\tilde{q}_{2} K_{2}(q'_{2},t'_{2};\tilde{q}_{2},t_{2}) \delta(\tilde{q}_{2}-q_{2}) G(q'_{1},t'_{1};q_{1},t_{1};\tilde{q}_{2},t_{2};q_{2},t_{2})  \nonumber \\
		&=K_{2}(q'_{2},t'_{2};q_{2},t_{2}) G(q'_{1},t'_{1};q_{1},t_{1};q_{2},t_{2};q_{2},t_{2}) 
		\;, \label{anotheritr}
	\end{align}
	}
	and the propagator for the upper corner path is given by
	\begin{align}
		K_{\ulcorner}\left(q'_{1},t'_1,q'_{2},t'_2;q_{1},t_{1},q_{2},t_{2}\right)&=\left\langle q'_{1},q'_{2}\left|U_{1}U_{2}\right|q_{1},q_{2}\right\rangle \nonumber \\
		&=\iint d\bar{q}_{1} d\bar{q}_{2} \left\langle q'_{1},q'_{2}\left|U_{1}\right|\bar{q}_{2},\bar{q}_{1}\right\rangle \left\langle\bar{q}_{2},\bar{q}_{1}\left|U_{2}\right|q_{1},q_{2}\right\rangle \nonumber \\
		&=\iint d\bar{q}_{1} d\bar{q}_{2} \left\langle q'_{1},q'_{2}\left|U_{1}\right|\bar{q}_{2},\bar{q}_{1}\right\rangle \left\langle \bar{q}_{1}|q_{1}\right\rangle \left\langle \bar{q}_{2} \left|U_{2}\right|q_{2}\right\rangle \nonumber \\
		&=\iint d\bar{q}_{1} d\bar{q}_{2} \left\langle q'_{1},q'_{2}\left|U_{1}\right|\bar{q}_{2},\bar{q}_{1}\right\rangle \delta(\bar{q}_{1}-q_{1}) K_{2}(\bar{q}_{2},t'_{2};q_{2},t_{2}) \nonumber \\
		&=\int d\bar{q}_{2} \left\langle q'_{2}\left|\left\langle q'_{1} \left|U_{1}\right|q_{1}\right\rangle\right|\bar{q}_{2}\right\rangle K_{2}(\bar{q}_{2},t'_{2};q_{2},t_{2}) \nonumber\\
		&=\int d\bar{q}_{2} \left\langle q'_{2}\left|G'(q'_{1},t'_{1};q_{1},t_{1};\hat{q_{2}})\right|\bar{q}_{2}\right\rangle K_{2}(\bar{q}_{2},t'_{2};q_{2},t_{2}) \nonumber\\
		&=\int d\bar{q}_{2} \left\langle q'_{2}|\bar{q}_{2}\right\rangle G'(q'_{1},t'_{1};q_{1},t_{1};q'_{2},t'_{2};\bar{q}_{2},t'_{2})  K_{2}(\bar{q}_{2},t'_{2};q_{2},t_{2}) \nonumber \\
		&=\int d\bar{q}_{2} \delta(q'_{2}-\bar{q}_{2}) G'(q'_{1},t'_{1};q_{1},t_{1};q'_{2},t'_{2};\bar{q}_{2},t'_{2})  K_{2}(\bar{q}_{2},t'_{2};q_{2},t_{2}) \nonumber \\
		&=G'(q'_{1},t'_{1};q_{1},t_{1};q'_{2},t'_{2};q'_{2},t'_{2})  K_{2}(q'_{2},t'_{2};q_{2},t_{2})
		\;. \label{anotheritr2}
	\end{align}
	Then we find that the propagators for the upper and lower paths are not the same. This implies that the quantum evolution of the system with interaction is path-dependent. Of course, this path-dependent feature is a direct consequence of the violation of the consistency condition \eqref{cptblt7}.
	\\\\
	\emph{Example}: Here we will show the explicit example. We choose to work with the following Lagrangians
	\begin{align}
		L_{1}&=\frac{m\dot{q_{1}}^{2}}{2}+kq_{1}q_{2} = \frac{m\dot{q_{1}}^{2}}{2}+Fq_{1}\;\;\;\;\;\; ; F=kq_{2}\nonumber \\
		L_{2}&=\frac{m\dot{q_{2}}^{2}}{2}
		\;. \label{L1L2}
	\end{align}
	The propagator for a free particle with the constant force $F$ with the Lagrangian $L=\frac{m\dot{q}^{2}}{2}+Fq$ is given by \cite{Feynman}
	\begin{align}
		K^{F}\left(q',t';q,t\right)&=\sqrt{\frac{m}{2\pi i  (t'-t)}} e^{i\left\{\frac{m}{2}\frac{(q'-q)^{2}}{t'-t}+\frac{F}{2}(q'+q)(t'-t)-\frac{F^{2}}{24m}(t'-t)^{3}\right\}}
		\;. \label{free+force}
	\end{align}
	We will proceed the same transition given in figure \ref{fig:mypath}. Then the propagator of the lower corner path can be written as
	\begin{align}
		K_{\lrcorner}\left(q'_{1},t'_1,q'_{2},t'_2;q_{1},t_{1},q_{2},t_{2}\right)&=\left\langle q'_{1},q'_{2}\left|U_{2}U_{1}\right|q_{1},q_{2}\right\rangle \nonumber \\
		&=\sqrt{\frac{m}{2\pi i  (t'_{2}-t_{2})}} e^{\frac{im}{2}\frac{(q'_{2}-q_{2})^{2}}{t'_{2}-t_{2}}}  \nonumber \\
		&\quad\cross\int d\tilde{q}_{2} \Bigg\langle \tilde{q}_{2}\Bigg|\sqrt{\frac{m}{2\pi i  (t'_{1}-t_{1})}} e^{i\left\{\frac{m}{2}\frac{(q'_{1}-q_{1})^{2}}{t'_{1}-t_{1}}+\frac{F}{2}(q'_{1}+q_{1})(t'_{1}-t_{1})-\frac{F^{2}}{24m}(t'_{1}-t_{1})^{3}\right\}} \Bigg|q_{2}\Bigg\rangle \nonumber \\
		&=\sqrt{\frac{m}{2\pi i  (t'_{2}-t_{2})}} e^{\frac{im}{2}\frac{(q'_{2}-q_{2})^{2}}{t'_{2}-t_{2}}} \quad\sqrt{\frac{m}{2\pi i  (t'_{1}-t_{1})}} e^{\frac{im}{2}\frac{(q'_{1}-q_{1})^{2}}{t'_{1}-t_{1}}}  \nonumber \\
		&\quad\cross\int d\tilde{q}_{2} \left\langle \tilde{q}_{2}\left|e^{i\left\{\frac{k\hat{q}_{2}}{2}(q'_{1}+q_{1})(t'_{1}-t_{1})-\frac{k^{2}\hat{q}^{2}_{2}}{24m}(t'_{1}-t_{1})^{3}\right\}} \right|q_{2}\right\rangle \nonumber \\
		&=K_2(q'_{2},t'_{2};q_{2},t_{2}) K_1(q'_{1},t'_{1};q_{1},t_{1})  \nonumber \\
		&\quad\cross \int d\tilde{q}_{2} \left\langle \tilde{q}_{2}|q_{2}\right\rangle e^{i\left\{\frac{k\tilde{q_{2}}}{2}(q'_{1}+q_{1})(t'_{1}-t_{1})-\frac{k^{2}\tilde{q_{2}}^{2}}{24m}(t'_{1}-t_{1})^{3}\right\}} \nonumber \\
		&=K_2(q'_{2},t'_{2};q_{2},t_{2}) K_1(q'_{1},t'_{1};q_{1},t_{1}) \nonumber \\
		&\quad\cross \int d\tilde{q}_{2} \delta(\tilde{q_{2}}-q_{2}) e^{i\left\{\frac{k\tilde{q_{2}}}{2}(q'_{1}+q_{1})(t'_{1}-t_{1})-\frac{k^{2}\tilde{q_{2}}^{2}}{24m}(t'_{1}-t_{1})^{3}\right\}}  \nonumber \\
		&=K_2(q'_{2},t'_{2};q_{2},t_{2}) K_1(q'_{1},t'_{1};q_{1},t_{1}) e^{i\left\{\frac{kq_{2}}{2}(q'_{1}+q_{1})(t'_{1}-t_{1})-\frac{k^{2}q^{2}_{2}}{24m}(t'_{1}-t_{1})^{3}\right\}}
		\;, \label{freeitr1}
	\end{align}
	We shall proceed with the same computation for the upper path and we obtain
	\begin{align}
		K_{\ulcorner}\left(q'_{1},t'_1,q'_{2},t'_2;q_{1},t_{1},q_{2},t_{2}\right)&=\left\langle q'_{1},q'_{2}\left|U_{1}U_{2}\right|q_{1},q_{2}\right\rangle \nonumber \\
		&=\int d\bar{q}_{2} \Bigg\langle q'_{2}\Bigg| \sqrt{\frac{m}{2\pi i  (t'_{1}-t_{1})}} e^{i\left\{\frac{m}{2}\frac{(q'_{1}-q_{1})^{2}}{t'_{1}-t_{1}}+\frac{F}{2}(q'_{1}+q_{1})(t'_{1}-t_{1})-\frac{F^{2}}{24m}(t'_{1}-t_{1})^{3}\right\}} \Bigg|\bar{q}_{2}\Bigg\rangle \nonumber \\ 
		&\quad\cross\sqrt{\frac{m}{2\pi i  (t'_{2}-t_{2})}} e^{\frac{im}{2}\frac{(q'_{2}-q_{2})^{2}}{t'_{2}-t_{2}}} \nonumber \\
		&=\sqrt{\frac{m}{2\pi i  (t'_{1}-t_{1})}} e^{\frac{im}{2}\frac{(q'_{1}-q_{1})^{2}}{t'_{1}-t_{1}}} \int d\bar{q}_{2} \left\langle q'_{2}\left|e^{i\left\{\frac{k\hat{q}_{2}}{2}(q'_{1}+q_{1})(t'_{1}-t_{1})-\frac{k^{2}\hat{q}^{2}_{2}}{24m}(t'_{1}-t_{1})^{3}\right\}} \right|\bar{q}_{2}\right\rangle \nonumber \\
		&\quad\cross\sqrt{\frac{m}{2\pi i  (t'_{2}-t_{2})}} e^{\frac{im}{2}\frac{(q'_{2}-q_{2})^{2}}{t'_{2}-t_{2}}} \nonumber \\
		&= K_1(q'_{1},t'_{1};q_{1},t_{1}) \int d\bar{q}_{2} \left\langle q'_{2}|\bar{q}_{2}\right\rangle e^{i\left\{\frac{k\bar{q_{2}}}{2}(q'_{1}+q_{1})(t'_{1}-t_{1})-\frac{k^{2}\bar{q_{2}}^{2}}{24m}(t'_{1}-t_{1})^{3}\right\}} \nonumber \\ 
		&\quad\cross K_2(q'_{2},t'_{2};q_{2},t_{2}) \nonumber \\
		&= K_1(q'_{1},t'_{1};q_{1},t_{1}) \int d\bar{q}_{2} \delta(q'_{2}-\bar{q}_{2}) e^{i\left\{\frac{k\bar{q_{2}}}{2}(q'_{1}+q_{1})(t'_{1}-t_{1})-\frac{k^{2}\bar{q_{2}}^{2}}{24m}(t'_{1}-t_{1})^{3}\right\}} \nonumber \\ 
		&\quad\cross K_2(q'_{2},t'_{2};q_{2},t_{2}) \nonumber \\
		&=K_1(q'_{1},t'_{1};q_{1},t_{1}) K_2(q'_{2},t'_{2};q_{2},t_{2}) e^{i\left\{\frac{kq'_{2}}{2}(q'_{1}+q_{1})(t'_{1}-t_{1})-\frac{k^{2}q'_{2}{}^{2}}{24m}(t'_{1}-t_{1})^{3}\right\}}   
		\;. \label{freeitr2}
	\end{align}
	This simple calculation shows that the interaction causes the violation of the relation \eqref{cptblt7} and consequently the commutation of the propagators. Of course, the path-independent is no longer applicable. In the geometry point of view, the present of the interaction can be viewed as a course of temporal space curvature and therefore, the parallel transport of different paths would give different results.
	
	\section{Conclusion}\label{section4}
	We succeed to capture the consistency condition for the multi-time evolution in terms of the Lagrangian as the consequence of the variation of the action on the space of time variables. This consistency condition implies that the action is invariant under the local deformation, fixing end-points, of the path on the space of time variables. Actually, if we think that the continuous path is constituted from tiny discrete elements, then, path-independent property in the continuous-time case is a direct consequence of path-independent in the discrete-time case. Furthermore, with this property, there is a family of paths(homotopy), sharing the endpoints, that can be continuously transformed to each other in $N$-dimensional space of time variables.
	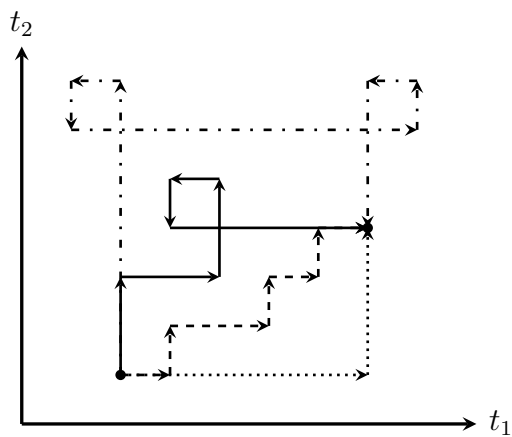
\begin{figure}[h]
		\centering
		\begin{tikzpicture}[scale=0.65]
			\foreach \coord/\label [count=\xi] in {
				{7.7,0.5}/{$t_{1}$},
				{-2,7.7}/{$t_{2}$}	
			}
			{
				\pgfmathsetmacro\anch{mod(\xi,2) ? "north" : "south"}
				\node[anchor=\anch] at (\coord) {\label};
			}
			\foreach \point in {(0,1),(5,4)}{
				\fill \point circle (3pt);
			}

			\draw[-stealth,line width=1.25pt] (-2,0) -- (7.2,0);
			\draw[-stealth,line width=1.25pt] (-2,0) -- (-2,7.7);
			
			\draw[-stealth,line width=1pt,loosely dashdotted] (0,1) -- (0,7);
			\draw[-stealth,line width=1pt,loosely dashdotted] (0,7) -- (-1,7);
			\draw[-stealth,line width=1pt,loosely dashdotted] (-1,7) -- (-1,6);
			\draw[-stealth,line width=1pt,loosely dashdotted] (-1,6) -- (6,6);
			\draw[-stealth,line width=1pt,loosely dashdotted] (6,6) -- (6,7);
			\draw[-stealth,line width=1pt,loosely dashdotted] (6,7) -- (5,7);
			\draw[-stealth,line width=1pt,loosely dashdotted] (5,7) -- (5,4);
			
			\draw[-stealth,line width=1pt] (0,1) -- (0,3);
			\draw[-stealth,line width=1pt] (0,3) -- (2,3);
			\draw[-stealth,line width=1pt] (2,3) -- (2,5);
			\draw[-stealth,line width=1pt] (2,5) -- (1,5);
			\draw[-stealth,line width=1pt] (1,5) -- (1,4);
			\draw[-stealth,line width=1pt] (1,4) -- (5,4);
			
			\draw[-stealth,line width=1pt,dashed] (0,1) -- (1,1);
			\draw[-stealth,line width=1pt,dashed] (1,1) -- (1,2);
			\draw[-stealth,line width=1pt,dashed] (1,2) -- (3,2);
			\draw[-stealth,line width=1pt,dashed] (3,2) -- (3,3);
			\draw[-stealth,line width=1pt,dashed] (3,3) -- (4,3);
			\draw[-stealth,line width=1pt,dashed] (4,3) -- (4,4);
			\draw[-stealth,line width=1pt,dashed] (4,4) -- (5,4);
			
			\draw[-stealth,line width=1pt,dotted] (0,1) -- (5,1);
			\draw[-stealth,line width=1pt,dotted] (5,1) -- (5,4);

		\end{tikzpicture}
		\caption{The possible paths, including shortest path, zigzag path and path with loops,  $t_{1}-t_{2}$ space from the initial point to the final point.}
		\label{fig:pathindependent}
	\end{figure}
	\\
	The consistency condition for the multi-time quantum evolution in terms of Feynman's path integrals is derived. The important point is the path-independent feature of the multi-time propagator which can be summarised as follows. In general, there are an infinite number of paths from the initial point to the final point on the space of time variables, see figure \ref{fig:pathindependent} in the case of two-time variables. With a set of Lagrangians $\{L_1,L_2,...,L_N \}$ satisfying the consistency condition, the propagator remains unchanged under the variation of the path on the space of time variables, and of course, this is nothing but the path-independent feature of the multi-time propagator. This would suggest that, apart from taking all possible paths in the configuration space as we normally do in the standard single-time path integration, one may need to take also the all possible paths in the space of time variables for the case of the multi-time path integration\footnote{This terminology arises also in the context of integrable systems \cite{Frank2}}. In the view of the geometry, the path-independent feature can be captured in terms of the parallel transport process on the flat space of time variables since the curvature vanishes. Then the consistency condition for a set of Lagrangians can be viewed as the zero curvature condition.
	\section{Appendix}
	Here we show the derivation of the transition that evolves from $(t_{1},t_{2})$ to $(\hat{t}_{1},\hat{t}_{2})$, see figure \ref{fig:loop2timeA}.
	\begin{align}
		\left\langle \hat{q}_{1},\hat{q}_{2}\big|\Phi(\hat{t}_{1},\hat{t}_{2})\right\rangle &= \left\langle \hat{q}_{1},\hat{q}_{2}\left|U'_{2}U'_{1}U_{2}U_{1}\right|\Phi(t_{1},t_{2})\right\rangle \nonumber  \\
		\Phi\left(\hat{q}_{1},\hat{q}_{2},\hat{t}_{1},\hat{t}_{2}\right)&=\iint dq_{1} dq_{2} \left\langle \hat{q}_{1},\hat{q}_{2}\left|U'_{2}U'_{1}U_{2}U_{1}\right|q_{1},q_{2}\right\rangle\left\langle q_{2},q_{1}\big|\Phi(t_{1},t_{2})\right\rangle \nonumber  \\
		&=\iint dq_{1} dq_{2} \left\langle \hat{q}_{1},\hat{q}_{2}\left|U'_{2}U'_{1}U_{2}U_{1}\right|q_{1},q_{2}\right\rangle\Phi(q_{1},q_{2},t_{1},t_{2}) \nonumber \\
		&=\iiiint d\tilde{q}_{1} d\tilde{q}_{2} dq_{1} dq_{2} \left\langle \hat{q}_{1},\hat{q}_{2}\left|U'_{2}U'_{1}U_{2}\right|\tilde{q}_{1},\tilde{q}_{2}\right\rangle \left\langle \tilde{q}_{2},\tilde{q}_{1}\left|U_{1} \right|q_{1},q_{2}\right\rangle\Phi(q_{1},q_{2},t_{1},t_{2}) \nonumber  \\
		&=\iiiint d\tilde{q}_{1} d\tilde{q}_{2} dq_{1} dq_{2} \left\langle \hat{q}_{1},\hat{q}_{2}\left|U'_{2}U'_{1}U_{2}\right|\tilde{q}_{1},\tilde{q}_{2}\right\rangle \left\langle \tilde{q}_{1}\left|U_{1}\right|q_{1}\right\rangle \left\langle \tilde{q}_{2}\big|q_{2}\right\rangle \Phi(q_{1},q_{2},t_{1},t_{2}) \nonumber  \\
		&=\iiiint d\tilde{q}_{1} d\tilde{q}_{2} dq_{1} dq_{2} \left\langle \hat{q}_{1},\hat{q}_{2}\left|U'_{2}U'_{1}U_{2}\right|\tilde{q}_{1},\tilde{q}_{2}\right\rangle \left\langle \tilde{q}_{1}\left|U_{1}\right|q_{1}\right\rangle \delta(\tilde{q}_{2}-q_{2}) \Phi(q_{1},q_{2},t_{1},t_{2}) \nonumber\\
		&=\iiiint\!\!\!\iint dq'_{1} dq'_{2} d\tilde{q}_{1} d\tilde{q}_{2} dq_{1} dq_{2} \left\langle \hat{q}_{1},\hat{q}_{2}\left|U'_{2}U'_{1}\right|q'_{1},q'_{2}\right\rangle \left\langle q'_{2},q'_{1}\left|U_{2}\right|\tilde{q}_{1},\tilde{q}_{2}\right\rangle \left\langle \tilde{q}_{1}\left|U_{1}\right|q_{1}\right\rangle \nonumber\\
		&\quad\cross  \delta(\tilde{q}_{2}-q_{2}) \Phi(q_{1},q_{2},t_{1},t_{2}) \nonumber  \\
		&=\iiiint\!\!\!\iint dq'_{1} dq'_{2} d\tilde{q}_{1} d\tilde{q}_{2} dq_{1} dq_{2} \left\langle \hat{q}_{1},\hat{q}_{2}\left|U'_{2}U'_{1}\right|q'_{1},q'_{2}\right\rangle \left\langle q'_{2}\left|U_{2}\right|\tilde{q}_{2}\right\rangle \left\langle q'_{1}\big|\tilde{q}_{1}\right\rangle \left\langle \tilde{q}_{1}\left|U_{1}\right|q_{1}\right\rangle \nonumber\\
		&\quad\cross \delta(\tilde{q}_{2}-q_{2}) \Phi(q_{1},q_{2},t_{1},t_{2}) \nonumber  \\
		&=\iiiint\!\!\!\iint dq'_{1} dq'_{2} d\tilde{q}_{1} d\tilde{q}_{2} dq_{1} dq_{2} \left\langle \hat{q}_{1},\hat{q}_{2}\left|U'_{2}U'_{1}\right|q'_{1},q'_{2}\right\rangle \left\langle q'_{2}\left|U_{2}\right|\tilde{q}_{2}\right\rangle \delta(q'_{1}-\tilde{q}_{1}) \left\langle \tilde{q}_{1}\left|U_{1}\right|q_{1}\right\rangle \nonumber\\
		&\quad\cross \delta(\tilde{q}_{2}-q_{2}) \Phi(q_{1},q_{2},t_{1},t_{2}) \nonumber  \\
		&=\iiiint\!\!\!\iiiint d\bar{q}_{1} d\bar{q}_{2} dq'_{1} dq'_{2} d\tilde{q}_{1} d\tilde{q}_{2} dq_{1} dq_{2} \left\langle \hat{q}_{1},\hat{q}_{2}\left|U'_{2}\right|\bar{q}_{1},\bar{q}_{2}\right\rangle \left\langle \bar{q}_{1},\bar{q}_{2} \left|U'_{1}\right|q'_{1},q'_{2}\right\rangle \left\langle q'_{2}\left|U_{2}\right|\tilde{q}_{2}\right\rangle  \nonumber\\
		&\quad\cross \delta(q'_{1}-\tilde{q}_{1}) \left\langle \tilde{q}_{1}\left|U_{1}\right|q_{1}\right\rangle \delta(\tilde{q}_{2}-q_{2}) \Phi(q_{1},q_{2},t_{1},t_{2}) \nonumber  \\
		&=\iiiint\!\!\!\iiiint d\bar{q}_{1} d\bar{q}_{2} dq'_{1} dq'_{2} d\tilde{q}_{1} d\tilde{q}_{2} dq_{1} dq_{2} \left\langle \hat{q}_{1},\hat{q}_{2}\left|U'_{2}\right|\bar{q}_{1},\bar{q}_{2}\right\rangle \left\langle \bar{q}_{1} \left|U'_{1}\right|q'_{1}\right\rangle \left\langle \bar{q}_{2}\big|q'_{2}\right\rangle \left\langle q'_{2}\left|U_{2}\right|\tilde{q}_{2}\right\rangle  \nonumber\\
		&\quad\cross \delta(q'_{1}-\tilde{q}_{1}) \left\langle \tilde{q}_{1}\left|U_{1}\right|q_{1}\right\rangle \delta(\tilde{q}_{2}-q_{2}) \Phi(q_{1},q_{2},t_{1},t_{2}) \nonumber  \\
		&=\iiiint\!\!\!\iiiint d\bar{q}_{1} d\bar{q}_{2} dq'_{1} dq'_{2} d\tilde{q}_{1} d\tilde{q}_{2} dq_{1} dq_{2} \left\langle \hat{q}_{1},\hat{q}_{2}\left|U'_{2}\right|\bar{q}_{1},\bar{q}_{2}\right\rangle \left\langle \bar{q}_{1} \left|U'_{1}\right|q'_{1}\right\rangle  \delta(\bar{q}_{2}-q'_{2}) \left\langle q'_{2}\left|U_{2}\right|\tilde{q}_{2}\right\rangle  \nonumber\\
		&\quad\cross \delta(q'_{1}-\tilde{q}_{1}) \left\langle \tilde{q}_{1}\left|U_{1}\right|q_{1}\right\rangle \delta(\tilde{q}_{2}-q_{2}) \Phi(q_{1},q_{2},t_{1},t_{2}) \nonumber  \\
		&=\iiiint\!\!\!\iiiint d\bar{q}_{1} d\bar{q}_{2} dq'_{1} dq'_{2} d\tilde{q}_{1} d\tilde{q}_{2} dq_{1} dq_{2} \left\langle \hat{q}_{2}\left|U'_{2}\right|\bar{q}_{2} \right\rangle  \left\langle \hat{q}_{1}\big|\bar{q}_{1}\right\rangle \left\langle \bar{q}_{1} \left|U'_{1}\right|q'_{1}\right\rangle \delta(\bar{q}_{2}-q'_{2}) \left\langle q'_{2}\left|U_{2}\right|\tilde{q}_{2}\right\rangle  \nonumber\\
		&\quad\cross \delta(q'_{1}-\tilde{q}_{1}) \left\langle \tilde{q}_{1}\left|U_{1}\right|q_{1}\right\rangle \delta(\tilde{q}_{2}-q_{2}) \Phi(q_{1},q_{2},t_{1},t_{2}) \nonumber \\
		&=\iiiint d\bar{q}_{2} d\tilde{q}_{1} dq_{1} dq_{2} \left\langle \hat{q}_{2}\left|U'_{2}\right|\bar{q}_{2}\right\rangle \int d\bar{q}_{1} \delta(\hat{q}_{1}-\bar{q}_{1})  \left\langle \bar{q}_{1} \left|U'_{1}\right|q'_{1}\right\rangle \int dq'_{2} \delta(\bar{q}_{2}-q'_{2}) \left\langle q'_{2}\left|U_{2}\right|\tilde{q}_{2}\right\rangle  \nonumber\\
		&\quad\cross \int dq'_{1}\delta(q'_{1}-\tilde{q}_{1}) \left\langle \tilde{q}_{1}\left|U_{1}\right|q_{1}\right\rangle \int d\tilde{q}_{2} \delta(\tilde{q}_{2}-q_{2}) \Phi(q_{1},q_{2},t_{1},t_{2}) \nonumber  \\
		&=\iiiint d\bar{q}_{2} d\tilde{q}_{1} dq_{1} dq_{2} \left\langle \hat{q}_{2}\left|U'_{2}\right|\bar{q}_{2}\right\rangle  \left\langle \bar{q}_{2}\left|U_{2}\right|q_{2}\right\rangle \left\langle \hat{q}_{1} \left|U'_{1}\right|\tilde{q}_{1}\right\rangle  \left\langle \tilde{q}_{1}\left|U_{1}\right|q_{1}\right\rangle \Phi(q_{1},q_{2},t_{1},t_{2}) \nonumber  \\
		&=\iint dq_{1} dq_{2} \int d\bar{q}_{2} \left\langle \hat{q}_{2}\left|U'_{2}(\hat{t}_{2}-t'_{2})\right|\bar{q}_{2}\right\rangle  \left\langle \bar{q}_{2}\left|U_{2}(t'_{2}-t_{2})\right|q_{2}\right\rangle \nonumber  \\
		&\quad\cross \int d\tilde{q}_{1} \left\langle \hat{q}_{1} \left|U'_{1}(\hat{t}_{1}-t'_{1})\right|\tilde{q}_{1}\right\rangle  \left\langle \tilde{q}_{1}\left|U_{1}(t'_{1}-t_{1})\right|q_{1}\right\rangle \Phi(q_{1},q_{2},t_{1},t_{2}) \nonumber  \\
		&=\iint dq_{1} dq_{2} \int d\bar{q}_{2} K_{2}(\hat{q}_{2},\hat{t}_{2};\bar{q}_{2},t'_{2}) K_{2}(\bar{q}_{2},t'_{2};q_{2},t_{2}) \int d\tilde{q}_{1} K_{1}(\hat{q}_{1},\hat{t}_{1};\tilde{q}_{1},t'_{1}) K_{1}(\tilde{q}_{1},t'_{1};q_{1},t_{1}) \nonumber  \\
		&\quad\cross \Phi(q_{1},q_{2},t_{1},t_{2})
		\;. \label{proveloop2timenewnew}
	\end{align}
	This proves the equation of transition \eqref{loop2timenewnew} on the evolution-counter clockwise of two-time system.
	
	
	\section*{Acknowledgements}
	We would like to thank Pichet Vanichchapongjaroen for the valuable discussion. S. Sungted is supported by Development and Promotion of Science and Technology Talents Project (DPST).
	\\
	\\
	Conflict of Interest: The authors declare that they have no
	conflicts of interest.
	

\end{document}